\documentclass{aa}  

\usepackage{graphicx}
\usepackage{txfonts}
\usepackage{lipsum}
\usepackage{subcaption}         
\usepackage{lscape}             
\usepackage{placeins}           

\usepackage[version=4]{mhchem} 
\usepackage{booktabs}
\usepackage{array}
\usepackage{comment}
                                
\usepackage{xcolor}
\usepackage[hidelinks]{hyperref}
\hypersetup{
    colorlinks=true,
    linkcolor=blue,
    citecolor=blue
}

\begin{document}

   \title{Kinetic Monte Carlo modeling of ammonium formate \ce{NH4+HCOO-} and ammonium cyanate \ce{NH4+OCN-} in interstellar ices}

   \subtitle{}

   \titlerunning{Kinetic Monte Carlo modeling of \ce{NH4+HCOO-} and \ce{NH4+OCN-} in interstellar ices}

%
%
%

\author{Joseph Salaris\inst{1}\corrauth{joseph.salaris@ru.nl} \and Robin Toonen\inst{1}\email{robin.toonen@ru.nl} \and Rens Waters\inst{2}\email{rens.waters@ru.nl} \and Herma Cuppen\inst{1}\email{herma.cuppen@ru.nl}}

\institute{Institute for Molecules and Materials, Radboud University, PO Box 9010, 6500 GL Nijmegen, The Netherlands
\and  Department of Astrophysics/IMAPP, Radboud University, PO Box 9010, 6500 GL Nijmegen, The Netherlands 
}

   \date{Received September 30, 20XX}
 
  \abstract
   {JWST observations have revealed evidence for \ce{NH4+} in the ices of the Chamaeleon I dark molecular cloud, with \ce{HCOO-} and \ce{OCN-} being likely counter ions. Laboratory experiments have demonstrated that ammonium formate (\ce{NH4+HCOO-}) and ammonium cyanate (\ce{NH4+OCN-}) can form through acid-base reactions involving \ce{NH3} at cryogenic temperatures. }
   {Here the formation of these two ammonium salts are simulated to investigate their formation in dark-cloud ices and assess their sensitivity to uncertainties in the chemical reaction network, including the recently considered HOCO-mediated pathway for \ce{CO2} formation, as well as variations in the cosmic-ray ionization rate, cloud collapse timescale and density, and elemental depletion.} 
   {Our modeling framework consists of microscopic kinetic Monte Carlo (KMC) simulations of grain-surface chemistry augmented by UCLCHEM gas-phase and physical cloud evolution simulations. The microscopic KMC method provides for explicit simulation of the location, movement and reaction of ice species, making it particularly well-suited to model acid-base chemistry of \ce{NH4+}-salt formation, occurring within the ice mantle over extended timescales.}
   {The simulations produce the canonical ice-evolution sequence, with an initially \ce{H2O}-rich ice followed by a layer dominated by CO and \ce{CO2}. The salt production is dominated by \ce{NH4+HCOO-}, which is found to form predominantly during the \ce{H2O}-rich stage as its production is limited by the availability of \ce{NH3}. \ce{NH4+OCN-} is found to form later at the interface between the \ce{H2O}- and CO-rich ice when the formation of its precursor acid HNCO is enabled. 
   }
   {Despite the variations explored, the total \ce{NH4+} abundance, particularly relative to \ce{H2O}, remains remarkably robust and consistent with the observed abundance. Depending on the variation, a redistribution between the counterions \ce{HCOO-} and \ce{OCN-} is observed.}


   \maketitle
\nolinenumbers

\section{Introduction}
Recent \emph{James Webb Space Telescope} (JWST) observations have affirmed the presence of ammonium salts in interstellar ices within dark molecular clouds \citep{McClure2023}. Ammonium salts are a class of chemical compounds containing the ammonium ion (\ce{NH4+}) and a counter anion, which can form through acid-base reactions between ammonia (\ce{NH3}) and acids. The \ce{NH4+} ion was detected toward two background sources, J110621 and NIR38, in the Chamaeleon I dense molecular cloud through its mid-infrared absorption feature at 6.85~$\mu$m. This absorption feature has long been observed in interstellar ices, initially primarily toward relatively warm young stellar objects in observations with the \emph{Infrared Space Observatory} (ISO) and later with \emph{Spitzer} \citep{Keane2001,Boogert2008}. The feature was proposed to arise in part from the $\nu_4$ vibrational mode of \ce{NH4+}, based on laboratory experiments \citep{Schutte2003}. Now the JWST observations provide strong evidence for the presence of \ce{NH4+} in the cold, dense molecular-cloud environment of Chamaeleon I \citep{McClure2023}. The inferred column density of \ce{NH4+} is of the same order as ammonia: $N(\ce{NH4+}) \approx N(\ce{NH3})$, indicating that ammonium is present in significant abundance.

The presence of ammonium in dark molecular clouds indicates that ammonium salts can already form at an early stage of star formation at very low temperatures ($T\sim 10$ K). In this stage, ices forming on nanosized carbon/silicate grains, regulate much of the chemistry and are crucial to explaining the rich molecular inventory found in the dense areas of the interstellar medium (ISM). In contrast to the gas -- which contains most of the mass of the molecular cloud -- the grain-surface chemistry enables the formation of molecules, such as \ce{CH3OH}, \ce{H2O} and \ce{NH3}, as well as more complex species (involving more atoms). Locked in the ices, these molecules are the main reservoir of volatile elements that influence the chemistry in subsequent stages and environments of star formation (\citealp[see, e.g.,][]{Quenard2018,Coutens2020,Rocha2025}).

Ammonium salts are typically more thermally stable than their neutral precursors, allowing them to remain on grain surfaces well above the desorption temperatures of their constituent molecules. For example, experiments \citep{Kruczkiewicz2021,Noble2012,Gerakines2024} have shown that ammonium formate (\ce{NH4+HCOO-}), ammonium acetate (\ce{NH4+CH3COO-}), and ammonium cyanide (\ce{NH4+CN-}) desorb at temperatures higher or comparable to the water-ice desorption temperature \citep{Fraser2001}. Ammonium salts may therefore retain nitrogen on grain surfaces after the desorption of their neutral precursors and other volatile ice constituents, potentially affecting the evolution of the elemental nitrogen budget through the various stages of star formation. For example, the study of ammonium formate and ammonium acetate desorption by \citet{Kruczkiewicz2021} showed that the ammonium sublimation zone is expected to lie within the water snowline, closer to the core of a protostar. This could alter the distribution of nitrogen across the protoplanetary disk relative to the conventional picture in which N-bearing species desorb with the water ice mantle, and would have important implications for identifying the main nitrogen carriers in the environment of a developing star. Since ammonium salts can play a role in the formation of complex organic molecules (COMs) and (pre)biotic species \citep{Puzzarini2024}, they can also act as reservoirs of molecules and functional groups that would otherwise desorb at lower temperatures, potentially preserving precursors for subsequent COM and prebiotic chemistry.

The implications of nitrogen retention in ammonium salts on grains extend beyond the protostellar environment and may also help explain the apparent nitrogen depletion in comets. While nitrogen is only lightly depleted in the ISM and appears to remain predominantly in the gas as diffuse clouds evolve into dense clouds \citep{Savage1996,Jenkins2009}, a significant depletion is observed in comets \citep{Filacchione2019}. Measurements by the Giotto and Rosetta missions toward comets 1P/Halley and 67P/Churyamov-Gerasimenko respectively, found N/C ratios up to an order of magnitude lower than solar-system values \citep{Jessberger1988,Fray2017,Lodders2010}, while the N/Si ratio in 1P/Halley was also found to be lower than the solar value \citep{Geiss1988}. Ammonium salts have emerged as a potential reservoir of this missing nitrogen, with their presence in comet 67P supported by several independent measurements, including the detection of sublimation products in the coma \citep{Altwegg2020}, identification of the 3.2~$\mu$m absorption band with \ce{NH4+} \citep{Quirico2016,Poch2020,Poch2023}, and direct \emph{in-situ} measurements of ammonium salts in dust grains released from the cometary nucleus \citep{Altwegg2022}. Although several formation scenarios have been proposed, the salts may at least in part have a prestellar origin \citep{Altwegg2022}. Correlations between protostellar and cometary chemical compositions \citep{Altwegg2019,Drozdovskaya2019,Rocha2024} further support the possibility that ammonium salts were formed on icy dust grains within the molecular cloud and subsequently incorporated into cometary material.

Given their importance and confirmed detection, it would be desirable to take \ce{NH4+} salts into account in astrochemical models of dark-cloud ices, elucidate their formation on grain surfaces and assess how it depends on physical and chemical conditions. However, their explicit formation has only recently been considered in astrochemical models; to our knowledge, only one modeling effort has so far investigated ammonium salt formation, focusing on the ammonium sulfur salt \ce{NH4+SH-} \citep{Vitorino2024}. The choice of which ammonium salts to include in such models remains open, since the detection of \ce{NH4+} does not uniquely identify the counterion with which it is associated in the ice. \ce{HCOO-} and \ce{OCN-} are particularly promising candidates, given their detection in interstellar ices and evidence for their association with \ce{NH4+} in laboratory and cometary material. Recent JWST observations toward starless and protostellar regions of molecular clouds \citep{McClure2023,Rocha2024,Rocha2025} have secured the attribution of the 4.62~$\mu$m band to \ce{OCN-} \citep[e.g.][]{Geballe1984,vanBroekhuizen2005}. The detection of \ce{HCOO-} is less straightforward, as its spectral signature lies in the COM fingerprint region (6.8--8.6~$\mu$m), where spectral features from multiple molecules overlap. Nevertheless, \ce{HCOO-} has been identified in JWST observations toward both low- and high-mass protostars \citep{Rocha2024}, where it was found to be the main carrier of the 7.2 and 7.4~$\mu$m bands. Moreover, direct evidence for ammonium formate comes from comet 67P, where a spectral reflectance feature of the cometary surface was found to match well with the laboratory spectrum of \ce{NH4+HCOO-} mixed with analogue cometary surface material \citep{Poch2020}.

In this work, we investigate the formation of ammonium formate (\ce{NH4+HCOO-}) and ammonium cyanate (\ce{NH4+OCN-}) in the ices of dark molecular clouds. Salt formation in interstellar ices has long been proposed to proceed through energetic processing of ice mantles followed by acid-base reactions (\citealp[e.g.][]{Grim1989}). Although irradiation is an effective laboratory means of producing acids and ions, it may not be required for salt formation under astrophysical conditions \citep{Schutte2003,vanBroekhuizen2005}. Indeed, ammonium salts can readily form through acid-base reactions with sufficiently low activation barriers to proceed at cryogenic temperatures \citep{Theule2013}. Quantum chemical calculations \citep{Park2006} and laboratory experiments have shown that ammonium ions and their counter ions can form upon co-deposition of \ce{NH3} with acids such as isocyanic acid (\ce{HNCO}) \citep{Novozamsky2001,Raunier2003,Mispelaer2012} and formic acid (\ce{HCOOH}) \citep{Schutte1999,Galvez2010,Bergner2016}. We therefore focus on the formation of \ce{NH4+HCOO-} and \ce{NH4+OCN-} through the following thermal acid-base reactions on grain surfaces:
\begin{align}
    \ce{NH3 + HCOOH &\to NH4+HCOO-}\label{reacAmmoniumFormate}\\
    \ce{NH3 + HNCO &\to NH4+OCN-}.\label{reacAmmoniumCyanate}
\end{align}
At low temperatures, these reactions are expected to be facilitated provided the reactants are positioned close to each other. \citeauthor{Bergner2016} and \citeauthor{Mispelaer2012} have shown that reactions ~(\ref{reacAmmoniumFormate}) and~(\ref{reacAmmoniumCyanate}) have relatively low activation barriers ($70$ and $48$ K respectively) under such conditions. This demonstrates that thermal acid-base chemistry can provide a viable pathway for ammonium salt formation if the corresponding acids are present in the ice. \ce{HCOOH} has been securely detected in interstellar ices toward protostars with JWST \citep{Rocha2024,Rocha2025}, while \ce{HNCO} has not yet been detected in interstellar ices; however, gas-phase abundances of \ce{HNCO} have been found to correlate with solid \ce{OCN-} \citep{Oberg2009}.

Our modeling framework combines a microscopic kinetic Monte Carlo (KMC) model of grain-surface chemistry with the rate-equation code UCLCHEM \citep{Holdship2017}, which follows the gas-phase chemistry and the physical evolution of a collapsing molecular cloud. Together, these models enable us to perform comprehensive gas- and grain-surface chemistry simulations during the evolution of a molecular cloud. The KMC program allows for a microscopic simulation where the location and movement of each individual atomic/molecule species is followed and each reaction event is explicitly simulated. These properties make the microscopic KMC method uniquely suited to model reactions between acid and base species. In particular, these reactions are expected to proceed rapidly once the reactants are in close proximity at low temperatures \citep{Mispelaer2012,Bergner2016}, making the spatial distribution and mobility of the reactants important for determining their overall efficiency. 

Rate-equation models, by contrast, provide a macroscopic, mean-field description of the underlying microscopic processes. The relatively recently developed non-diffusive chemistry formalism (\citealp[see][]{Ruaud2015,Chang2016,Shingledecker2018,Jin2020}) can account for reactions that do not require prior thermal diffusion and may therefore provide a more appropriate treatment of these acid-base reactions than conventional Langmuir-Hinshelwood diffusion reactions. However, non-diffusive chemistry methods have primarily been developed to treat reactions with radicals leading to the formation of COMs, and their applicability to salt-formation reactions involving abundant and stable ice species, such as \ce{NH3}, remains to be firmly established.
The microscopic KMC treatment therefore provides a more direct approach for investigating ammonium-salt formation under these conditions.

We use this modeling framework to assess the sensitivity of ammonium formate and ammonium cyanate formation within the ice to uncertainties in the chemical reaction network and in the physical and chemical conditions of the molecular cloud. 
In particular, we investigate the effect of the \ce{CO2} formation network, including the recently considered \ce{HOCO} radical pathway, on ammonium salt formation. Furthermore, we explore the sensitivity of ammonium salt formation to variations in the cosmic-ray ionization rate, the timescale and final density of cloud collapse, and the assumed level of elemental depletion.
This paper is structured as follows. Section~\ref{secMethods} describes the modeling framework and Section~\ref{subsecIceEvo} presents the baseline model and its resulting ice evolution and ammonium salt formation. We investigate variations in the chemical reaction network in Sect.~\ref{subsecVaryChem} and in the physical conditions in Sect.~\ref{subsecVaryPhys}. The results of the chemical and physical variations are synthesized in Sect.~\ref{subsecNH4+formation}, where we also compare the modeled ice composition and ammonium-salt abundances with observations. Conclusions are presented in Sect.~\ref{secConclusions}.

\section{Modeling method}\label{secMethods}
The grain surface chemistry was simulated using a ``continuous-time random-walk kinetic Monte Carlo'' algorithm, which is an implementation of the microscopic kinetic Monte Carlo (KMC) technique that keeps track of the positions of the species on the grain. For a detailed description, an overview of the theoretical background and its application to surface astrochemistry, see \citet{Cuppen2013}. The KMC code used is an adaptation of the one applied in \citet{Cuppen2007, Cuppen2009, Lamberts2014, Simons2020,Jimenez2025}.

In particular, a similar setup to \citet{Jimenez2025} was used, where the gas-phase chemistry was simulated using the rate-equation code UCLCHEM (version 3.2) which includes freeze-out, \ce{H2} formation and a few basic hydride formation reactions as part of its grain-surface chemistry. The obtained abundances, densities, and photon fluxes were then used as input to the KMC simulation, from which the adsorption fluxes were calculated using standard kinetic gas theory (\citealp[see, e.g.,][]{Cuppen2017}), as well as the photodissociation rates. 
We note that this leads to a one-way coupling of the gas and grain chemistry. In both UCLCHEM and the KMC, thermal and chemical desorption is simulated and our setup may lead to inconsistencies if the desorption rates between the two models differ significantly. In that case, the adsorption fluxes calculated from the gas abundances (simulated by UCLCHEM) would not reflect the desorption of grain species (simulated by the KMC). We assume this effect to be to be negligible at the low temperatures considered here. A post-analysis confirmed that feedback from desorption has a negligible effect on the gas-phase composition, with the exception of H and \ce{H2}.

The surface formation of \ce{H2} in UCLCHEM was found to be more efficient than observed in the KMC simulations, which affects the atomic H abundance in the gas. \ce{H2} formation through recombination on the grain is modeled by UCLCHEM using two formalisms: an effective kinetic rate equation treatment \citep{Cazaux2002,Cazaux2004} and Langmuir-Hinschelwood reactions. The use of these two processes independently from each other can lead to ``double'' formation of \ce{H2}, which is especially problematic for our one-way coupled setup between UCLCHEM and KMC simulations, as the atomic hydrogen is lost from the system. Moreover, although the \ce{H2}-formation efficiency implemented in the kinetic rate equation formalism is based on surface-model studies parameterized for a wide range of astrophysical conditions following \citet{Cazaux2002,Cazaux2004}, it does not consider a grain surface network where many reactions compete for atomic H, as is the case in our KMC simulations. This becomes important at higher densities where the fluxes of O, CO, and N onto the grain is of similar magnitude as atomic H. Here a significant fraction of the H landing onto the grain gets locked up in species such as \ce{H2O}, \ce{CH4}, and \ce{CH3OH} and also the routes for \ce{H2} formation change from \ce{H + H \to H2} to \ce{H2} abstraction reactions \citep{Simons2020}. 

In order to alleviate the above-mentioned issues within our model setup, we adopted an iterative strategy where the \ce{H2} formation efficiency employed in UCLCHEM for the kinetic rate equation treatment was reparametrized using \ce{H2} formation data from the KMC calculations. The new gas-phase data obtained in this way was then used again to rerun the KMC simulations. See Appendix~\ref{AppenH2formationEfficiency} for the details and implications of using the adjusted versus the default UCLCHEM \ce{H2}-formation efficiency. We note that in the most recent version of UCLCHEM ($\geq 4.0$) double formation of \ce{H2} has been addressed using a temperature-dependent switch between the two formalisms.

As KMC simulations can quickly become computationally expensive, we have kept the selection of transferred gas-phase abundances to a minimum, i.e., to atomic and molecular species that represent the main reservoirs of the elements \ce{H, C, O, and N} in the gas phase and/or are directly involved in the formation of the main ice components and ammonium salts: H, CO, O, C, N, \ce{H2, O2, and N2}.
The gas-phase chemistry evolves on much longer timescales (of the order of thousands of years) than the grain-surface chemistry considered in the KMC simulations (ranging from microseconds to days). Accordingly, the gas abundances, densities, and photon fluxes are passed from UCLCHEM to the KMC program at intervals of 1000 years. Between these updates, the KMC simulations evolve using a variable timestep, explicitly resolving the faster grain-surface processes.

The grain surface is represented by a $50\times 50$ lattice with periodic boundary conditions, onto which species can accrete and form layers. Spherical grains with a radius of $0.1$~$\mathrm{\mu m}$ and a site density of $10^{15}$~cm$^{-2}$ are assumed. Desorption and diffusive hopping are treated as thermally-activated Arrhenius-type processes, with activation energies being site-specific and determined by the binding energies of the species and their interactions with neighbouring particles. See \citet{Cuppen2007} for a detailed account of the calculation of the rate coefficients for these processes, where we have used a factor $0.4$ instead of $0.5$ for the ratio between diffusion and binding energy. The non-thermal processes included are chemical (reactive) desorption and dissociation, the latter being caused either through direct ultraviolet- or cosmic-ray induced photon radiation. Equations for the corresponding rate coefficients can be found in \citet{Cuppen2007, Simons2020}.

\subsection{Physical and Chemical Parameters}\label{subsecPhysicalParameters}
In our simulations of ice chemistry, we consider the evolution of a typical molecular cloud from its diffuse to dense stage. Our setup is closely aligned with that of \citet{Jimenez2025} and we consider the model of J110621 as a reference and example of a typical dense cloud. 
It is constructed as follows. We start with a ``Phase 0'' (UCLCHEM only), modeling the cloud as a static diffuse cloud with a H nuclei density of $n_\mathrm{H}=2\times10^3$~cm$^{-3}$, gas and dust temperature of $15$ K and a visual extinction of $A_\mathrm{v}=1.5$~mag.
In the next phase, ``Phase 1'', the cloud collapse is simulated by increasing the density ($n_\mathrm{H}$) towards a final density, here set to $n_\mathrm{H}=2\times10^5$~cm$^{-3}$. The collapse follows the free-fall treatment \citep{Brown1988, Rawlings1992,Holdship2017}, and is engineered to span a duration of $t_\mathrm{collapse}=10^5$~yr. Correspondingly, we let the visual extinction increase with density towards a set final value of $A_\mathrm{v}=47$~mag \citep{Jimenez2018, Bohlin1978, Frerking1982}
\begin{align}
    A_\mathrm{v}=A_\mathrm{v0}+\frac{n_{\ce{H}}\times r
    _\mathrm{out}}{1.6\times 10^{21}},
\end{align}
where $A_\mathrm{v0}$ is the initial value for the extinction ($1.5$ mag) and $r_\mathrm{out}$ is the radius of the cloud. Finally, ``Phase 2'' is simulated; a static period of $2\times 10^4$~yr under the same dense cloud conditions reached at the end of the collapse. For phase 0, we have only obtained gas-phase abundances, whereas for phases 1 and 2, both gas-phase (using UCLCHEM) and grain-surface abundances (with the KMC program) are obtained. A schematic overview of our modeling setup is provided in Appendix~\ref{AppenModelDetails} and Table~\ref{TabParamRef} reports the relevant parameter values used for our reference model. Please note that the temperature remains constant throughout the simulation. 
\begin{table}
\caption{Parameter values for the reference model.}\label{TabParamRef}
\centering
\begin{tabular}{lccc}
\hline\hline
    & \textbf{Phase 0} & \textbf{Phase 1} & \textbf{Phase 2} \\
    & static & collapse & static \\
\hline
Parameter &  &  &  \\
$n_{\ce{C}}/n_{\ce{H}}$\tablefootmark{a}      
                             & $2.23\times10^{-4}$ & $2.23\times10^{-4}$ & $2.23\times10^{-4}$ \\
$n_{\ce{O}}/n_{\ce{H}}$\tablefootmark{a}      
                             & $5.62\times10^{-4}$ & $5.62\times10^{-4}$ & $5.62\times10^{-4}$ \\
$n_{\ce{N}}/n_{\ce{H}}$\tablefootmark{a}      
                             & $6.18\times10^{-5}$ & $6.18\times10^{-5}$ & $6.18\times10^{-5}$ \\
$\zeta$ [s$^{-1}$]           & $5.2\times10^{-17}$ & $5.2\times10^{-17}$ & $5.2\times10^{-17}$ \\
$t_\mathrm{duration}$ [yr]   & $10^6$              & $10^5$              & $2\times10^4$       \\
$n_{\mathrm{H}}$ [cm$^{-3}$] & $2\times10^3$       & $2\times10^5$\tablefootmark{b}       & $2\times10^5$       \\
$T$ [K]                      & $15$                & $10$                & $10$                \\
$A_{\mathrm{v}}$ [mag]       & $1.5$               & $47$\tablefootmark{b}                & $47$                \\
\hline
\end{tabular}
\tablefoot{
    \tablefoottext{a}{Elemental abundances set from \citet{Jenkins2009}, minimally depleted case, $F_*=0$.}
    \tablefoottext{b}{Value at the end of the phase.}
}
\end{table}

The cosmic-ray ionization rate, here denoted by $\zeta$, is set to $5.2\times 10^{-17}$~s$^{-1}$, consistent with estimates based on \ce{H3+} observations \citep{Dalgarno2006, Obolentseva2024}. This value is around four times higher than the canonical value of $1.3\times 10^{-17}$~s$^{-1}$ -- commonly used in astrochemical codes. The elemental abundances can also have a strong effect on the final ice composition in our simulations. Since these are provided as initial gas abundances in phase 0, i.e., a young diffuse cloud, we have chosen to use abundances corresponding to minimal depletion as determined by \citet{Jenkins2009} ($F_*=0$ case). The elemental abundances of C, O, and N are listed in Table~\ref{TabParamRef}.

In Sect.~\ref{subsecVaryPhys}, we investigate variations in some of the key parameters listed in Table~\ref{TabParamRef} and discuss the resulting changes relative to the reference model. The effects of varying $\zeta$, including adopting the canonical value, are explored in Sect.~\ref{susbecVaryCRIR}. Other choices of elemental abundances are discussed in Sect.~\ref{subsecElemAbun}. Finally, the effects of adopting longer collapse times ($t_\mathrm{collapse}$) and higher final densities ($n_\mathrm{H}$) are examined in Sect.~\ref{subsecSlowerCollapse}.


\subsection{Reaction network}
The used reaction network builds on previously established grain-surface reaction schemes, with reaction pathways confirmed by laboratory experiments or quantum mechanical calculations. It combines the water formation network \citep{Lamberts2014}, the CO hydrogenation network \citep{Simons2020}, and the glycine formation network \citep{Ioppolo2020} used in earlier studies. We have extended the network to include the formation of \ce{NH4+HCOO-} and \ce{NH4+OCN-} through thermal acid–base reactions (\ref{reacAmmoniumFormate}) and (\ref{reacAmmoniumCyanate}), respectively. In 
addition, we include acid formation reactions, \ce{CO} oxidation pathways involving the intermediate \ce{HOCO}, and other \ce{CO2} formation pathways through \ce{O}-atom addition. An overview of all newly-added reactions is shown in Table \ref{TabAddedReactions}.

Reaction~(\ref{reacAmmoniumFormate}) was experimentally shown to be feasible under cold interstellar ice conditions \citep{Bergner2016}. We adopt the measured rate constant, which follows an Arrhenius expression with an activation barrier of $70$ K and a pre-exponential factor of $1.4\times 10^{-3}$ s$^{-1}$. Similarly, the rate for reaction~(\ref{reacAmmoniumCyanate}) is adopted from \citet{Mispelaer2012} (barrier of $48$ K, pre-exponential factor of $3.5\times 10^{-3}$ s$^{-1}$).
For both acid-base reactions, the kinetics were found to be described by two distinct processes: a relatively slow process attributed to the movement of the reactants, involving diffusion and/or changes in their relative orientation, and a fast process corresponding to the chemical reaction itself \citep{Bergner2016,Mispelaer2012}. We adopt only the ``pure'' chemical reaction rate constants of the latter process from these studies, as diffusion and the resulting meeting rate between reactants are explicitly treated in the KMC model. The reaction rate constant is thus applied whenever the acid and base occupy neighboring lattice sites.

The role of the \ce{HOCO} radical is central to the different model variations discussed in this work. Traditionally, astrochemical models have attributed \ce{CO2} formation primarily to the direct reaction:
\begin{align}
\ce{CO + OH \to CO2 + H}. \label{reacCO2direct}
\end{align}
This is supported by several laboratory studies \citep{Oba2010, Ioppolo2011, Noble2011} and is the basis for our variation, Model~B. However, quantum-mechanical \citep{Goumans2008, Molpeceres2023}, molecular dynamics \citep{Arasa2013}, and recent experimental work \citep{Qasim2019, Ishibashi2024} suggest that \ce{HOCO} can be stabilized in the ice, leading to \ce{CO2} via a two-step process:
\begin{align}
    \ce{CO + OH &\to HOCO}\label{reacHOCO}\\
    \ce{HOCO + H &\to CO2 + H2}.\label{reacCO2adapt}
\end{align}
In our canonical model (Model~A), we adopt this mediated pathway. We also include the competing H-addition channel:
\begin{align}
\ce{HOCO + H &\to HCOOH}.\label{reacHCOOHfromHOCO}
\end{align}
Due to a lack of quantitative branching ratios, we assume equal branching among the three possible outcomes of the \ce{HOCO + H} reaction (forming \ce{CO2}, \ce{HCOOH}, or \ce{CO + H2O}). The rate for reaction~(\ref{reacHOCO}) is adopted from the \ce{H2O}-ice calculations of \citet{Molpeceres2023}, using a value of approximately $10^{10}$ s$^{-1}$. For consistency in comparing Model~A and Model~B (see Sect.~\ref{subsecEffectCO2}), the same rate constant is used for reaction~(\ref{reacCO2direct}) as for reaction~(\ref{reacHOCO}).

The formation of formic acid (\ce{HCOOH}) is treated via three pathways. First, we include the radical-radical reaction \ce{HCO + CO} \citep{Hudson1999, Bennett2010B} and a competing channel \ce{HCO + OH \to CO + H2O} with an equal branching ratio \citep{Dontgen2016}. Second, we include the barrierless reaction \ce{CH2OH + O \to HCOOH + H}, alongside an alternative channel to \ce{H2CO + OH} with equal branching ratio, as excess thermal energy may cause well-skipping in this product \citep{Seetula1994}. Third, we include the hydrogenation of the hydrocarboxyl radical (\ce{HOCO}), which laboratory studies suggest is a primary route to \ce{HCOOH} without external energetic processing \citep{Ioppolo2010}.

Finally, we consider the reaction of \ce{HOCO} with nitrogen atoms to enable the formation of \ce{HNCO}:
\begin{align}
\ce{HOCO + N \to HOCON,}\label{reacHOCON}
\end{align}
followed by a series of barrierless hydrogenation reactions (see Table \ref{TabAddedReactions}). In Sect.~\ref{subsecVaryHNCO}, we compare this route to an alternative direct pathway
\begin{align}
\ce{CO + NH \to HNCO,}\label{reacHNCOstraight}
\end{align}
which is omitted from our standard models due to the large uncertainties in its rate coefficient.

Destruction pathways for the two ammonium salts are not included in the reaction network due to lack of experimentally constrained dissociation rates and the uncertain nature of their destruction under interstellar conditions. While direct photodissociation is expected to be inefficient given the stable ionic nature of these salts, energetic-particle processing may induce destruction through ionization-driven processes, as briefly discussed in Sect.~\ref{subsecNH4+formation}.
Dissociation reactions for all other species are included and rates are taken where available from \citet{Heays2017}. Otherwise, canonical values are used. 

With the UCLCHEM simulations we have used the gas-phase chemical network from the UMIST 2012 database \citep{McElroy2013} and the default grain network (based on \citealp{Quenard2018}).

\section{Results and discussion}

\subsection{Ice evolution}\label{subsecIceEvo}
\begin{figure*}[h!]
   \sidecaption
   \includegraphics[width=12.97cm]{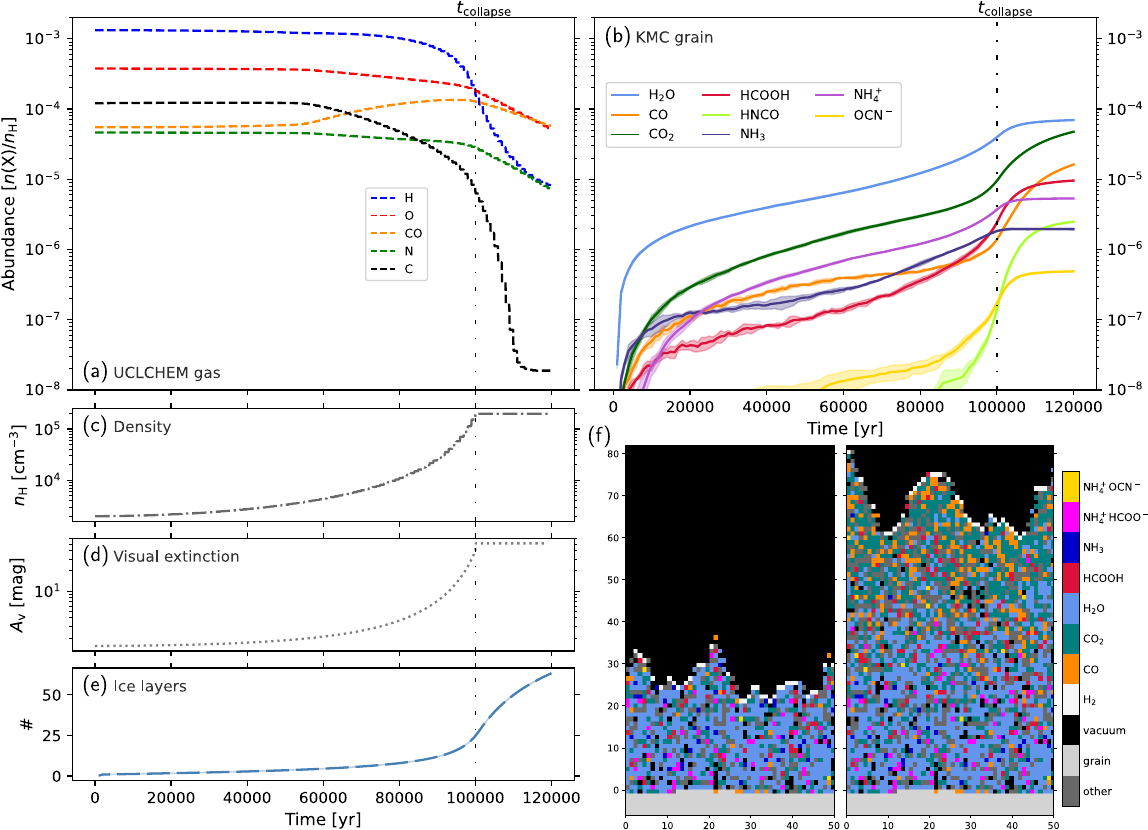}
      \caption{Time evolution of selected gas-phase (a) and grain-surface (b) species abundances during collapse and dense cloud conditions (Phase~1 and~2). The corresponding evolution of the cloud density (c), visual extinction (d), and number of ice monolayers (e) is shown in the lower-left panels. 
      Side cross-sections of the ice mantles are shown in panel~(f), at the end of Phase~1 (left), before severe CO freeze-out; and at the end of Phase~2 (right), after CO freeze-out. 
      Each pixel represents an atom or molecule of a given species, as indicated by the legend. 
      Gas-phase chemistry and cloud evolution (density and visual extinction) are modeled with UCLCHEM, from which gas abundances, density and visual extinction are provided to the KMC program simulating the grain-surface ice chemistry.
      KMC abundance profiles and the number of ice layers show mean and standard deviation (shaded area) based on 4 independent runs.}\label{testFigAggIceEvo}
\end{figure*}

The time-dependent cloud parameters provided to the KMC grain-surface simulation consist of the gas-phase abundances, cloud density, and visual extinction calculated with UCLCHEM. A selection of the gas abundances, together with the cloud density and visual extinction during Phase~1 and~2 of the simulation is shown in panels~(a), (c) and~(d) of Fig.~\ref{testFigAggIceEvo} respectively. Panel~(b) shows the time evolution of abundances of several ice species resulting from the KMC simulation. The corresponding growth of the ice mantle in terms of the number of monolayers is shown in panel~(e).
Here, and throughout the remainder of this work, time is reset to $t=0$ at the beginning of Phase~1 and $t_\mathrm{collapse}$ denotes the end of Phase~1 (start of Phase~2); the Phase-0 evolution is not shown. 

The cloud collapse model results in a rapid increase in density towards the end of Phase~1 ($t \gtrsim 7\times10^4$~yr), until dense cloud conditions are reached at $t_\mathrm{collapse}=10^5$~yr. In the subsequent Phase~2, the density remains constant at $n_\mathrm{H}=2\times10^5$~cm$^{-3}$ and the ice mantle continues to grow significantly. The ice thickness roughly doubles in size from $\sim30$ to $\sim60$ monolayers within a span of $20,000$ years after $t_\mathrm{collapse}$, driven by the increase in accretion flux and enhanced retention of species on the grain at high density and visual extinction.

The grain-surface abundances follow the canonical ice evolution scenario \citep{Oberg2011}, where ice formation is broadly divided into two stages that occur before and after significant depletion of \ce{CO} from the gas phase while freezing out onto the grain. Prior to \ce{CO} freeze-out, a water-rich ice mantle (polar ice) forms through the efficient hydrogenation of atoms and simple molecules, leading to the formation of hydrides such as \ce{H2O} and \ce{NH3}. This stage is followed by the formation of a \ce{CO}-rich ice (apolar ice), dominated by \ce{CO2} in addition to frozen \ce{CO}. In our model, the collapse time, $t_\mathrm{collapse}=10^5$~yr, approximately marks the transition between these two regimes: during Phase~1 we observe steady production of \ce{H2O} and \ce{NH3}, whereas in Phase~2 the abundances of these hydrides level off while the \ce{CO2} abundance increases rapidly. In our model, the onset of significant \ce{CO} freeze-out in Phase~2 is accompanied by a decrease in the gas-phase abundance of atomic H, reducing its availability for grain-surface chemistry and limiting the continued hydrogenation of accreted species. This contributes to the leveling off of the \ce{H2O} and \ce{NH3} abundances.

Panel~(f) of Fig.~\ref{testFigAggIceEvo} shows two cross-sections of the ice mantle, the left is at the end of Phase~1 and the right at the end of Phase~2. These cross-sections confirm the formation of two ice fractions: a polar layer dominated by blue representing \ce{H2O} and an apolar layer dominated by orange and teal, representing \ce{CO} and \ce{CO2} respectively. In both images, the ice mantle is covered by a single \ce{H2} layer. Molecular hydrogen does not accumulate because the binding energy of \ce{H2} to itself is too low to support multilayer growth (\citealp[see, e.g.,][]{Lee1972,Cuppen2007}), a feature that is explicitly accounted for in our KMC model.

We find that \ce{NH4+} salts predominantly form during the early stage of ice formation (Phase~1) and that their production is limited by the availability of \ce{NH3}. 
Beyond $t_\mathrm{collapse}$, the abundance of \ce{NH4+} increases only slightly following the cessation of \ce{NH3} production. In contrast, the formation of formic acid (\ce{HCOOH}) continues well into Phase~2, while the abundance of cyanic acid (\ce{HNCO}) only reaches appreciable levels during this later stage. The main formation channel for both acids is linked to the \ce{CO}-to-\ce{CO2} conversion chemistry through \ce{HOCO} (see Section~\ref{subsecEffectCO2}), leading to enhanced production with increasing levels of \ce{CO} abundance on the grain. This effect, combined with the reduced availability of \ce{NH3} for salt formation, leads to a significant increase in the abundances of both acids, especially at the interface between the two ice phases as can be seen in panel (f). 
Zooming in on the salt complexes, \ce{NH4+HCOO-} (magenta pixels) is found to be present throughout the \ce{H2O}-ice together with \ce{NH3}. The \ce{NH4+OCN-} complexes (gold pixels) are confined to upper layers of the \ce{H2O}-dominated ice, consistent with the delayed production of \ce{HNCO}. 
The near absence of \ce{NH4+} salts in the upper half of the ice is due to the absence of \ce{NH3} in this layer. 
In the following, we will refer to the model studied above as Model~A, which we regard as our baseline model. In the next sections, we consider variations of Model~A in which individual changes are made to the reaction network (Sect.~\ref{subsecVaryChem}) or to the input chemical and physical parameters (Sect.~\ref{subsecVaryPhys}).

\subsection{Changes to the chemical reaction network}\label{subsecVaryChem}

\subsubsection{Effect of inclusion of HOCO in the \ce{CO2} reaction network}\label{subsecEffectCO2}
\begin{figure*}
   \centering
   \includegraphics[width=\textwidth]{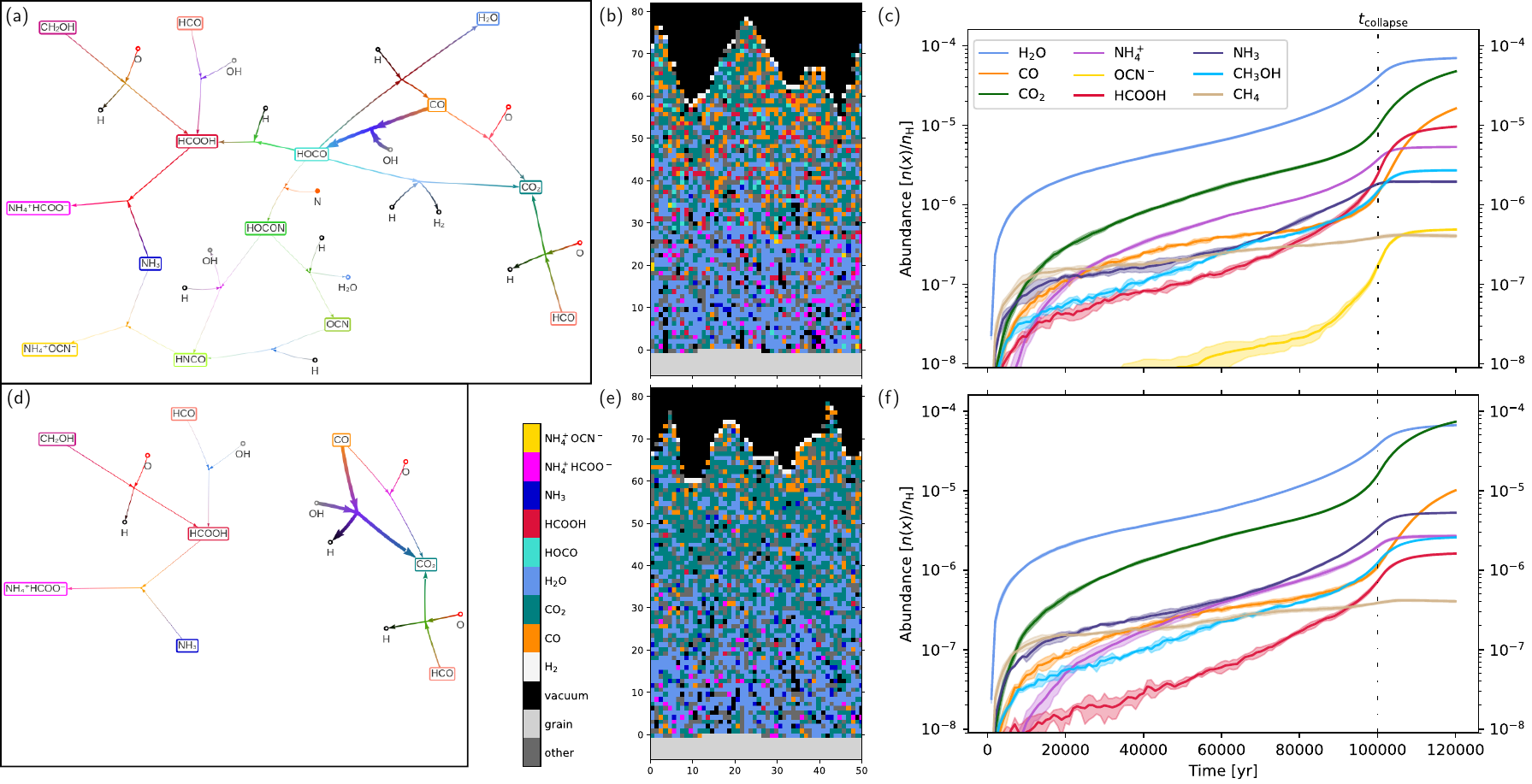}
      \caption{Comparison between Model~A, which includes the HOCO radical, and Model~B, in which \ce{CO2} is formed directly from \ce{CO + OH}. The top panels show the reaction network fluxes of the \ce{CO2} and \ce{NH4+} salt formation network (a), the ice cross-section at the end of Phase~2 (b), and the abundance profiles (c) for Model~A, while the corresponding results for Model~B are shown in the bottom panels (d-f). In the reaction network flux diagrams, the thickness of the arrows indicates the relative frequency of the reaction, and only reactions contributing more than 2\% to the formation of a species are shown. The abundance profiles show the mean and standard deviation (shaded area) based on four independent runs.}\label{FigCO2effect}
\end{figure*}

In our main model, Model~A, we have adopted the HOCO-mediated \ce{CO2} formation pathway and the associated HOCO reaction pathways. Here we investigate the effect of this HOCO-mediated treatment on the production of ammonium salts (\ce{NH4+HCOO-} and \ce{NH4+OCN-}) and on the overall ice composition by comparing Model~A with Model~B, in which HOCO is omitted from the reaction network and \ce{CO2} is formed through the direct pathway of reaction~(\ref{reacCO2direct}). 
A comparison between the two models is presented in Fig.~\ref{FigCO2effect}.

The reaction fluxes shown in Fig.~\ref{FigCO2effect}a illustrate how the \ce{CO2} formation flux connects, through \ce{HOCO}, to the formation pathways of \ce{NH4+HCOO-} and \ce{NH4+OCN-}. The hydrogen-addition reaction leading to \ce{CO2} (reaction~(\ref{reacCO2adapt})) competes with the other two channels, leading to \ce{H2O} and \ce{HCOOH}. Upon encounter with atomic H, the dominant reaction pathway likely depends on the relative orientation of H and HOCO. We adopt equal branching ratios between the three pathways, assuming no preferential orientation (see Table~\ref{TabAddedReactions}). This results in approximately equal formation rates of \ce{H2O}, \ce{CO2}, and \ce{HCOOH} from HOCO. In addition, HOCO can react with atomic nitrogen to form the intermediate HOCON, which can subsequently lead to the formation of the acid \ce{HNCO}. Both \ce{HCOOH} and \ce{HNCO} can react with \ce{NH3} in the ice through the acid–base reactions~(\ref{reacAmmoniumFormate}) and (\ref{reacAmmoniumCyanate}).

In contrast, in Model~B, the \ce{CO2} formation network is disconnected from the salt-production network, as illustrated in Fig.~\ref{FigCO2effect}d. The formation of \ce{NH4+HCOO-} proceeds through alternative pathways involving the methanol-formation network (via \ce{CH2OH}) and radical chemistry through HCO, largely independent of \ce{CO2} formation, which occurs directly via \ce{CO + OH}. No alternative route to the HOCO/HOCON pathway for the formation of \ce{HNCO} is included, and consequently no \ce{OCN-} is produced in this model without HOCO.

Figures~\ref{FigCO2effect}b and \ref{FigCO2effect}e show a comparison between Models~A and~B in terms of their ice cross-sections, while the abundance profiles of ice species throughout Phases~1 and~2 are compared in Figs.~\ref{FigCO2effect}c and \ref{FigCO2effect}f. Both models produce a CO-rich ice layer on top of a \ce{H2O}-dominated ice. The primary effect of including HOCO is a redistribution of the products of \ce{CO} oxidation from \ce{CO2} towards \ce{HCOOH}, with a corresponding increase in \ce{NH4+HCOO-}. This is evident from the increased presence of these species in the ice cross-sections and their higher final abundances. The HOCO-including model yields an approximately sixfold increase in the final relative abundance of \ce{HCOOH} ([\ce{HCOOH}]~$\sim 9.5 \times 10^{-6}$ for Model~A versus $\sim 1.6 \times 10^{-6}$ for Model~B) and a slightly more than twofold increase in \ce{NH4+HCOO-} ([\ce{NH4+HCOO-}]~$\sim 4.8 \times 10^{-6}$ for Model~A versus $\sim 2.7 \times 10^{-6}$ for Model~B). Furthermore, in Model~A, \ce{HOCO} is concentrated primarily in the upper ice layers, similar to \ce{HCOOH}. This reflects the reduced availability of \ce{NH3} for salt formation together with the relatively short duration of the CO-rich Phase~2, leaving species such as HCOOH and HOCO less efficiently converted.

The time evolution of the abundances of \ce{CO}, \ce{CH3OH}, and \ce{CH4} is largely unaffected by the inclusion of HOCO, as their abundance profiles are nearly identical in Models~A and~B. The similar \ce{CO} abundance is expected because \ce{CO} is consumed efficiently in both models: in Model~A, a fraction of \ce{CO} reacts with \ce{OH} to form \ce{HOCO}, whereas in Model~B the same reaction is treated as direct \ce{CO2} formation. Since the abundances of \ce{CH3OH} and \ce{CH4} are largely determined by the hydrogenation of CO and its hydrogenated intermediates, their abundance, in turn, also remains the same in both models. 
In contrast, the abundance of \ce{NH3} is strongly influenced: Model~B exhibits a final \ce{NH3} abundance approximately twice that of Model~A, whereas the \ce{NH4+} abundance is correspondingly lower. As discussed above, HOCO enhances the formation of acid species, leading to a decrease in \ce{NH3} as it reacts to form \ce{NH4+} salts.

We recommend using a reaction network where \ce{CO2} is formed through HOCO instead of the direct \ce{CO + OH} reaction, since this is consistent with several studies as indicated earlier \citep{Goumans2008,Molpeceres2023,Arasa2013,Qasim2019,Ishibashi2024}. Moreover, it results in a more realistic \ce{H2O}:\ce{CO2} ratio: in Model~A, \ce{H2O} remains the most abundant ice species throughout phases~1 and~2, whereas the more efficient conversion of \ce{CO} to \ce{CO2} in Model~B leads to a final \ce{CO2} abundance that exceeds that of \ce{H2O}.

Finally, we note that reaction~(\ref{reacHOCO}) is assigned a fixed rate coefficient based on calculations for a \ce{H2O}-rich ice environment \citep{Molpeceres2023}. This rate constant was found to vary substantially with the assumed ice matrix; a substantially higher rate coefficient was obtained when the reaction was modeled in a \ce{H2O} cluster than in a \ce{CO} cluster because water ice is much more efficient in dissipating excess energy \citep{Molpeceres2023}.  
Unfortunately, our model does not accommodate this change in rate coefficient because of change in local environment (yet) and a single value is set for the entire simulation. The rate coefficient used in Model~A is more relevant for Phase~1 ($t<t_\mathrm{collapse}$), when the bulk of the water ice is formed. During Phase~2 ($t_\mathrm{collapse}<t$), when the CO-rich ice layer emerges, the formation of \ce{HOCO} and species formed from \ce{HOCO} are likely overestimated. Indeed, we see that if we run the Model-A network using the rate coefficient relevant for a CO environment \citep{Molpeceres2023} for the entire run that \ce{HCOOH} and \ce{CO2} become less abundant, especially in Phase 2. 

The formation of \ce{NH4+} salts is likely largely unaffected by the switch in rate constant as these species form primarily during the early, water-rich stage. Moreover, their formation is limited by the availability of \ce{NH3}. We found that the abundance of \ce{NH4+} remained very similar for the two Model-A networks with different rate coefficients; $n(\ce{NH4+})/n_H$ values of $\sim 5\times 10^{-6}$ and $\sim 4\times 10^{-6}$ were observed at the end of Phase 2 for the \ce{H2O}-ice and CO-ice rate constants, respectively.

\subsubsection{Inclusion of reaction \ce{CO + NH \to HNCO}}\label{subsecVaryHNCO}
Model~A includes only one formation pathway for \ce{HNCO}, namely reaction~(\ref{reacHOCON}).
Here, we compare Model~A with a model that additionally includes the direct reaction between CO and the NH radical, i.e., reaction~(\ref{reacHNCOstraight}). We refer to this modified model as Model~C. 
\begin{figure}
   \centering
   \includegraphics[width=0.9\hsize]{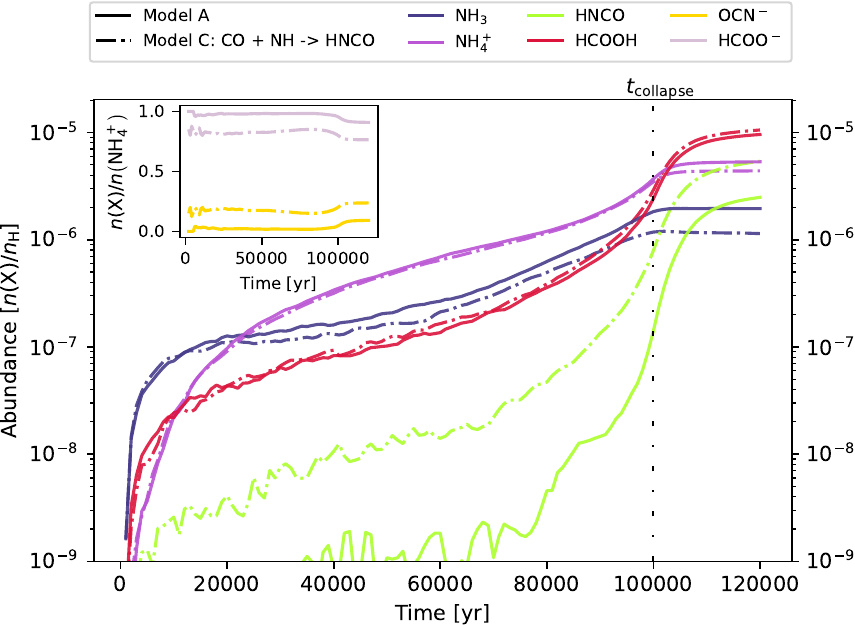}
      \caption{Abundance profiles during Phases~1 and~2 of cloud evolution for selected species relevant to \ce{NH4+} salt formation are shown for Models~A and~C. The latter model includes reaction~(\ref{reacHNCOstraight}) and exhibits enhanced \ce{HNCO} formation, with atomic N being diverted from \ce{NH3} and \ce{NH4+} towards \ce{HNCO}. The inset compares the abundance fraction of the two \ce{NH4+X-} salts (X$=$\ce{HCOO-} or \ce{OCN-}) relative to the total \ce{NH4+} abundance as a function of time between the two models.}\label{FigVaryReaction}
\end{figure}

Quantum chemical calculations, combined with experiments on \ce{HNCO} formation in 12~K Ar matrices \citep{Himmel2002}, suggest an activation energy of $4210$~K for reaction~(\ref{reacHNCOstraight}), rendering it negligible at temperatures of $10$~K. However, another experimental study aimed at reproducing astrochemically relevant conditions identified reaction~(\ref{reacHNCOstraight}) as the dominant formation pathway for \ce{HNCO}, compared to related reactions such as \ce{NH2 + CO \to HNCO + H} and \ce{H + NCO \to HNCO} \citep{Fedoseev2014}.
The rate coefficient for reaction~(\ref{reacHNCOstraight}) remains uncertain. Nevertheless \citeauthor{Fedoseev2014} suggested that the rate of \ce{H + CO \to HCO} provides an upper limit, since no \ce{H2CO} formation was detected in their experiments. Following the energy barriers reported by \citet{Fuchs2009} for \ce{H + CO \to HCO}, we use a rate of $2\times 10^{-5}$ s$^{-1}$ (corresponding to an energy barrier of $\sim 375$~K) for reaction~(\ref{reacHNCOstraight}).

A comparison of the ice abundance profiles during phases~1 and~2 of cloud evolution for a selection of species relevant to \ce{NH4+} salt formation is presented in Fig.~\ref{FigVaryReaction}. As expected, the inclusion of reaction~(\ref{reacHNCOstraight}) leads to an increase in the solid \ce{HNCO} abundance. Whereas for Model~A, the \ce{HNCO} fractional abundance remains low ($\sim10^{-9}$) during the early stage of ice evolution ($t < 80000$~yr), Model~C yields an \ce{HNCO} abundance already of the order of $10^{-8}$ through much of that time period. In Phase~2, the \ce{HNCO} abundance increases to values comparable to those of \ce{NH4+} ($4\times10^{-6}$), while for Model~A, the abundance reaches values comparable to that of \ce{NH3}.

In both models, peak production of \ce{HNCO} occurs near the end of Phase~1/start of Phase~2, corresponding to the onset of CO freeze-out. At this stage, CO becomes increasingly abundant, while hydrogenation of N towards \ce{NH} also remains efficient. 
The \ce{NH} radical can either react with H to form \ce{NH3} or with CO to form HNCO. With the rapid change in relative gas-phase abundance $n(\ce{H})/n(\ce{CO})$, the relative ice abundance of \ce{NH3} and HNCO also moves towards a preference of HNCO towards late times. The lower abundances of \ce{NH3} and \ce{NH4+} observed during Phase~2 in Model~C compared with Model~A can therefore be attributed to the allocation of nitrogen towards \ce{HNCO}. Concurrently, the \ce{HCOOH} abundance is nearly identical for both models. Thus, we can conclude that the CO-based chemistry involving \ce{HOCO} which forms \ce{HCOOH} is largely unaffected by the inclusion of reaction~(\ref{reacHNCOstraight}).

Finally, although the absolute abundance of \ce{NH4+} decreases slightly in Model~C, the dominant effect is a redistribution of the \ce{NH4+} salts among the counterions \ce{HCOO-} and \ce{OCN-}. The inset of Fig.~\ref{FigVaryReaction} shows the fraction of the two \ce{NH4+X-} species abundance relative to the total \ce{NH4+} abundance as a function of time. In Model~A, ammonium cyanate accounts for $\sim 2$--$5\%$ of the total \ce{NH4+} salt abundance, whereas in Model~C the \ce{NH4+} salt population is partitioned approximately $20:80$ between ammonium cyanate and ammonium formate. Hence, we conclude that the inclusion of reaction~(\ref{reacHNCOstraight}) leads to a significant increase in \ce{OCN-} formation. 

\subsection{Variation of physical parameters}\label{subsecVaryPhys}

\subsubsection{Effect of the cosmic-ray ionization rate}\label{susbecVaryCRIR}
Cosmic rays are a fundamental source of ionization within the interstellar medium (ISM), influencing both the gas-phase and grain-surface chemistry in molecular clouds.
The magnitude of the cosmic-ray ionization rate (CRIR) has remained uncertain for more than half a century. Directly estimated from the primary ionization rate of \ce{H2}, namely $\zeta$, this rate has been found to vary with the physical environment, ranging from values of the order of $10^{-18}$~s$^{-1}$ in well-shielded dense regions to $\sim10^{-15}$~s$^{-1}$ in diffuse clouds \citep{Spitzer1968,Dalgarno2006,Indriolo2012,Padovani2020}.
In addition, inferred values depend on the observational tracer and detection method used. A recent JWST detection of cosmic-ray-excited rovibrational \ce{H2} emission yielded $\zeta=1.7\times10^{-16}$~s$^{-1}$ \citep{Bialy2026}, whereas the canonical value adopted in many astrochemical models of dense clouds is $1.3\times10^{-17}$~s$^{-1}$ (\citealp[see, e.g.,][]{Garrod2013,Holdship2017,Iqbal2018,Jimenez2025}).

In our reference model, Model~A, we adopt $\zeta_A=5.2\times10^{-17}$~s$^{-1}$, following the estimates of \citet{Dalgarno2006,Obolentseva2024}. This value is four times the canonical dense-cloud value and remains well within the range inferred for dense molecular clouds. To investigate the sensitivity of grain-surface chemistry, and in particular the formation of \ce{NH4+} salts, to the cosmic-ray ionization rate, we additionally consider two models with lower and higher ionization rates: Model~D1 with $\frac{1}{4}\zeta_A=1.3\times10^{-17}$~s$^{-1}$ and Model~D2 with $2\zeta_A=1.04\times10^{-16}$~s$^{-1}$.

We identify two principal effects of varying the CRIR on the formation of \ce{NH4+} salts. The first is an increase in the production of atomic hydrogen. In our models, atomic H is produced in the gas phase through cosmic-ray ionization of \ce{H2} and subsequent ion-molecule chemistry, as computed by UCLCHEM. An increase in $\zeta$ therefore results in a larger flux of atomic H onto the grain surface, enhancing the hydrogenation of surface species. Most importantly, this leads to an increased production of \ce{NH3}, which enables the acid-base chemistry responsible for the formation of \ce{NH4+}.

\begin{figure}
   \centering
   \includegraphics[width=0.9\hsize]{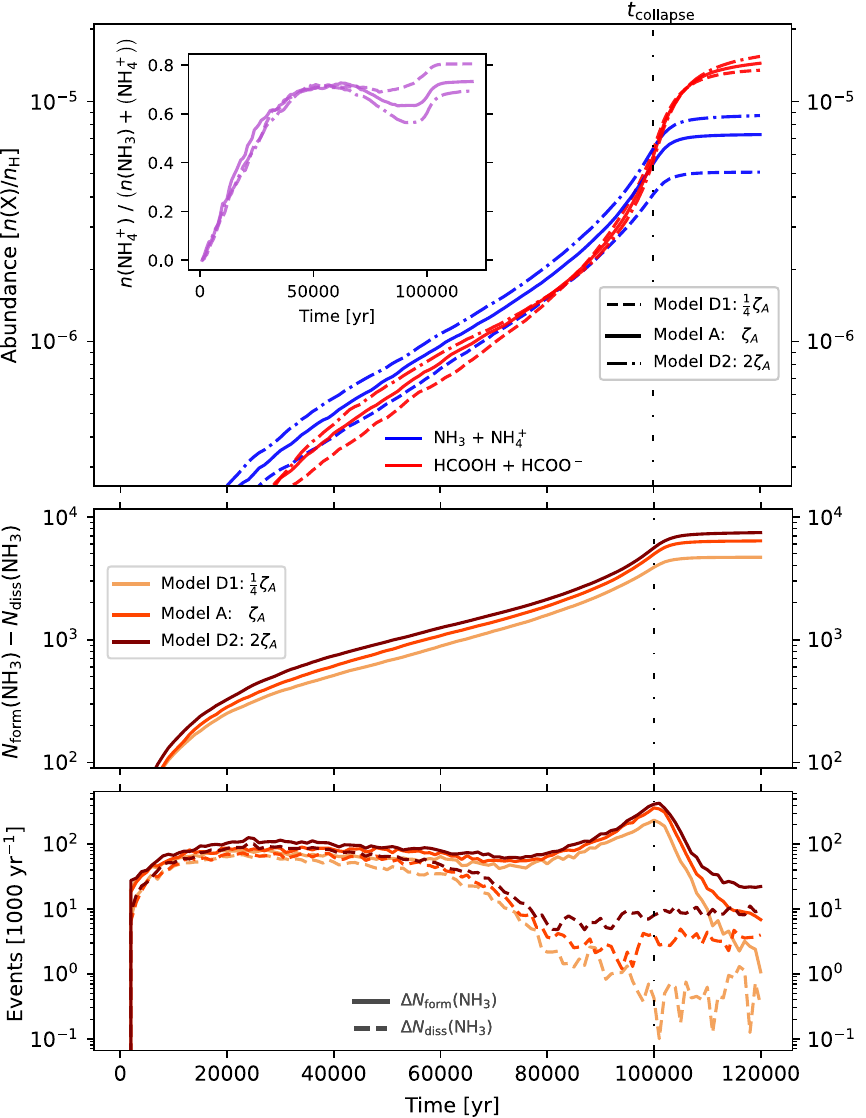}
    \caption{Comparison between models with different cosmic-ray ionization rates, $\zeta$, showing the total ammonia and formic acid reservoirs, $n(\ce{NH3})+n(\ce{NH4+})$ and $n(\ce{HCOOH})+n(\ce{HCOO-})$ respectively, during Phases~1 and~2 of the cloud evolution (panel~a). The inset shows the ratio $n(\ce{NH4+})/(n(\ce{NH3})+n(\ce{NH4+}))$. The net cumulative production of \ce{NH3} (panel~b) and the averaged binned frequencies of \ce{NH3} formation and dissociation events (panel~c) over the corresponding time period are also shown.}\label{FigVaryCRIR}
\end{figure}

The second effect is the enhanced photodissociation of ice species through the increased flux of cosmic-ray-induced UV photons. The photodissociation rates of the salts \ce{NH4+HCOO-} and \ce{NH4+OCN-}, however, remain unknown and are therefore not included in the present model. Moreover, salts are expected to be relatively resistant to UV-induced dissociation because of the strong electrostatic interaction between their constituent ions. This is further discussed in Section~\ref{subsecNH4+formation}.

For Model~A, the gas-phase abundance of atomic H is approximately $1.5\times10^{-3}$~cm$^{-3}$ throughout Phase~1 of the cloud evolution (see Sect.~\ref{subsecIceEvo}, Fig.~\ref{testFigAggIceEvo}a). 
Relative to Model~A, Model~D2 exhibits an increase of approximately $30\%$ in the atomic H abundance, whereas Model~D1 shows a decrease of approximately $40\%$. Similar increases and decreases are observed in the ice abundances of both \ce{NH3} and \ce{NH4+}, as shown in Fig.~\ref{FigVaryCRIR}a where the time evolution of the sum, $n(\ce{NH3})+n(\ce{NH4+})$, is plotted for each model. As shown in the inset, the fraction $n(\ce{NH4+})/(n(\ce{NH3})+n(\ce{NH4+}))$ decreases with increasing $\zeta$, particularly during Phase~2. This indicates that the enhanced production of \ce{NH3} outpaces its conversion into \ce{NH4+}, resulting in a larger reservoir of unreacted ammonia. Furthermore, these results demonstrate that, within our models, the CRIR effectively regulates the supply of atomic H available for hydrogenation of species on the grain. Because the formation of \ce{NH4+} salts is limited by the availability of \ce{NH3} (Sect.~\ref{subsecIceEvo}), increasing $\zeta$ enhances the production of \ce{NH3} and subsequently the abundance of \ce{NH4+}.

In contrast, varying the CRIR has little influence on CO-based chemistry or the formation of \ce{HCOOH} via the \ce{HOCO} intermediate. Instead, increasing $\zeta$ promotes the consumption of \ce{HCOOH} to form \ce{NH4^+HCOO^-}. As shown by the nearly constant total abundance of $n(\ce{HCOOH}) + n(\ce{HCOO^-})$ in Fig.~\ref{FigVaryCRIR}a, the CRIR primarily redistributes carbon and oxygen between the neutral acid and its ammonium salt rather than altering the overall \ce{HCOOH} budget.

Although a higher CRIR enhances the photodissociation of ice species, it paradoxically leads to a substantial increase in \ce{NH3} abundance. This indicates that enhanced production, driven by increased atomic \ce{H} abundance, outweighs the accompanying increase in photodissociation.
This competition is illustrated in panels~b and~c of Fig.~\ref{FigVaryCRIR}, where the net cumulative production of \ce{NH3} -- defined as the difference between formation via \ce{NH2 + H \to NH3} and total photodissociation -- increases with $\zeta$ throughout both phases of cloud evolution. Binned frequency analysis (Fig.~\ref{FigVaryCRIR}c) confirms that while a higher $\zeta$ increases the \ce{NH3} photodissociation, it more significantly enhances the formation rate due to the increased availability of surface atomic \ce{H}. Since formation exceeds dissociation across all models -- and this gap widens during late-stage cloud collapse ($8.0\times10^4~\mathrm{yr}<t<t_\mathrm{collapse}$) -- increased CRIR  increases the \ce{NH3} reservoir, ultimately facilitating more efficient \ce{NH4^+} salt formation.

\subsubsection{Collapse time and final cloud density}\label{subsecSlowerCollapse}
In our main model, Model~A, it is assumed that the molecular cloud collapses during Phase~1 from an initial hydrogen nuclei density of $n_\mathrm{H}=2\times10^3$~cm$^{-3}$ to a final density of $n_\mathrm{H}=2\times10^{5}$~cm$^{-3}$ over a timescale of $t_\mathrm{collapse}=10^5$~yr. 
The adopted collapse time is shorter than typically expected for the physical evolution of molecular clouds and may influence the grain-surface chemistry and the resulting ice composition. 
Using the parameterized collapse formalism of \citet{Rawlings1992}, the default retardation factor yields a collapse time of the order of 1~Myr for the diffuse-to-dense cloud evolution considered here.

\begin{figure*}[h!]
   \sidecaption
   \includegraphics[width=12.97cm]{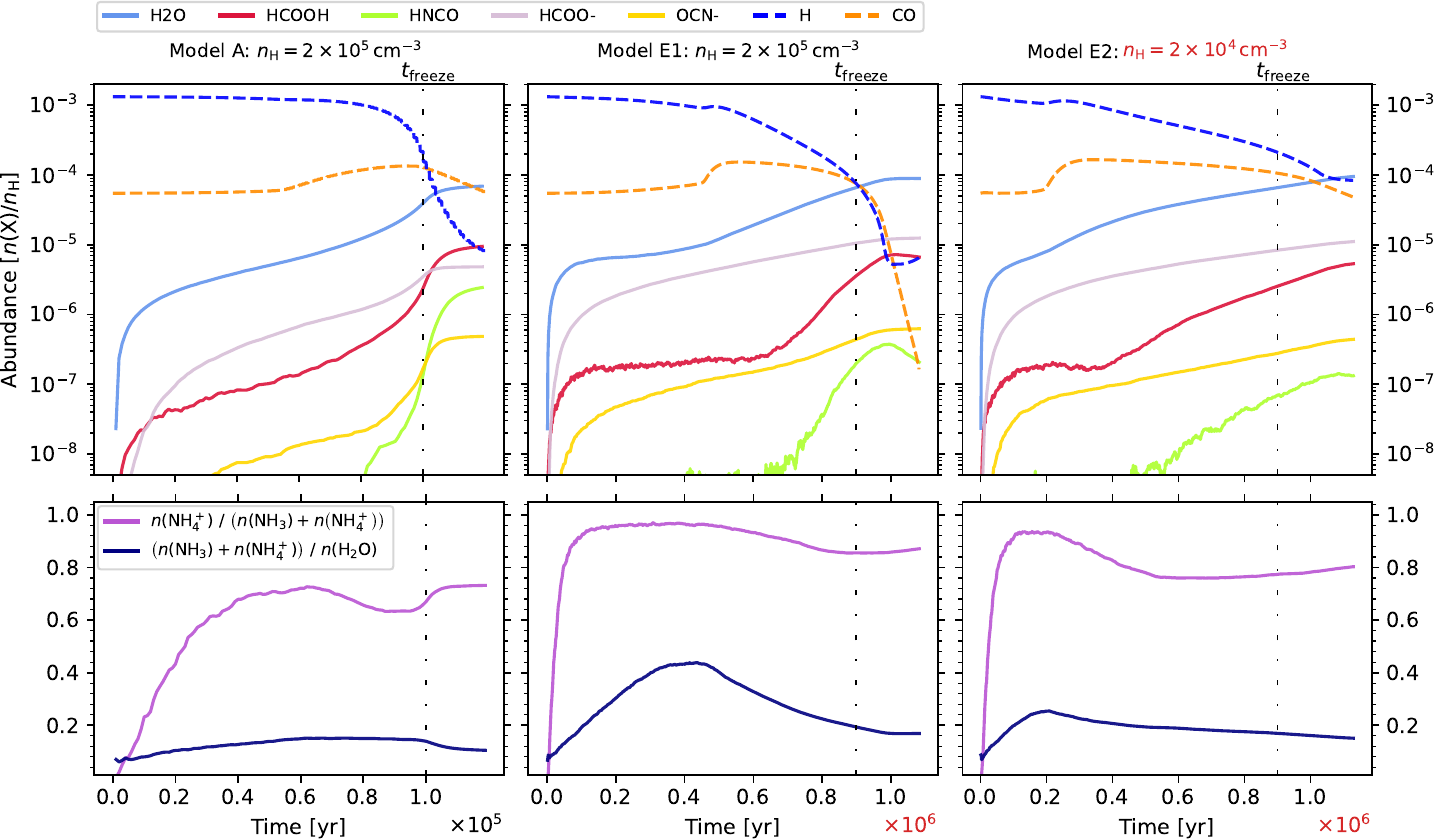}
      \caption{Comparison between models with different collapse times and final density, ordered from left to right according to increasingly slower cloud evolution. Values of the collapse timescale and final density that differ from those of Model~A are highlighted in red.
      Top panels: selection of gas-phase (dashed) and grain-surface (solid) abundance profiles for Models~A, E1, and~E2. Compared to Model~A, the slower collapse in Models~E1 and~E2 results in a more gradual freeze-out of CO and a less rapid decline in the gas abundance of atomic H. Bottom panels: fraction of \ce{NH4+} with respect to the total ammonia reservoir (\ce{NH4 + NH4+}) and the abundance of the total ammonia reservoir relative to \ce{H2O} over the corresponding time period.
      }\label{FigVaryDensityCollapseTime}
\end{figure*}

To assess the effect of the adopted cloud evolution on the grain-surface chemistry, and in particular the influence of a slower collapse on \ce{NH4+} salt formation, we compare Model~A with two models, Model~E1 and~E2, representing slower collapse. Model~E1 adopts a collapse time of $10^6$ years while retaining the same final density of $2\times10^5$~cm$^{-3}$. Model~E2 adopts the same collapse time but reaches a lower final density of $2\times10^4$~cm$^{-3}$. 
For both Models~E1 and~E2 the timespan of Phase~2 is set to $10^5$ years.

A selection of the resulting gas-phase and grain-surface abundance profiles is presented in the top panels of Fig.~\ref{FigVaryDensityCollapseTime}, ordered from the most rapid to the slowest cloud evolution (Models~A, E1, and E2). In the slower-collapse models, the freeze-out of gas species occurs more gradually, over several hundred thousand years rather than the tens of thousands of years in Model~A. Although significant CO freeze-out occurs later in absolute time in Models~E1 and~E2, it begins at an earlier stage of the density evolution, i.e., at a lower cloud density. While the onset of significant CO freeze-out coincides with the start of Phase~2 in Model~A, a comparable level of CO freeze-out is reached earlier relative to the collapse timescale in Models~E1 and~E2, at approximately $0.9\times t_\mathrm{collapse}=0.9\times10^6$~yr, indicated by $t_\mathrm{freeze}$ in Fig.~\ref{FigVaryDensityCollapseTime}.
The transition between a \ce{H2O}- and a \ce{CO}-rich ice is therefore roughly demarcated by $t_\mathrm{freeze}$, although in Models~E1 and~E2 this transition is much less abrupt and less well-defined. In both models, the gas-phase atomic H abundance stays comparable to the gas-phase CO abundance; consequently hydrogenation reactions, in particular \ce{H2O} formation, continue as CO increasingly dominates the ice composition. 
Meanwhile, the CO depletion found in Phase~2 of Model~E1 is extensive, but consistent with predictions from other astrochemical model simulations for similar timescale and density \citep{Jimenez2025}. In Model~E2, the longer timescale is compensated by the lower final density, resulting in a level of CO depletion comparable to that of Model~A instead of Model~E1.

Compared to Model~A, the formation of salt precursors is more gradual in the slower-collapse models and both \ce{NH4+HCOO-} and \ce{NH4+OCN-} are formed efficiently in Models~E1 and~E2; by $t=0.4\times t_\mathrm{collapse}$, these salts reach abundances that Model~A only achieves towards the end of its collapse. This efficiency is further evidenced by the higher ratios of \ce{NH4+} relative to the total ammonia reservoir (\ce{NH3 + NH4+}) seen in Models~E1 and~E2 and shown in the bottom panels of Fig.~\ref{FigVaryDensityCollapseTime}.

As also shown, at the end of Phase~2, all models have reached a similar \ce{NH3 + NH4+}/\ce{H2O} ratio. However, while in Model~A this ratio is rather constant during the simulation, the slow-collapse models have a clear peak in the ratio of the ammonia reservoir to \ce{H2O}. This is effect is most pronounced for Model~E1 around  $0.4\times10^6$~yr. It appears to coincide with the change in H and CO abundance in the gas phase, but the exact mechanism remains unclear.

\begin{figure*}[h!]
   \sidecaption
   \includegraphics[width=12.97cm]{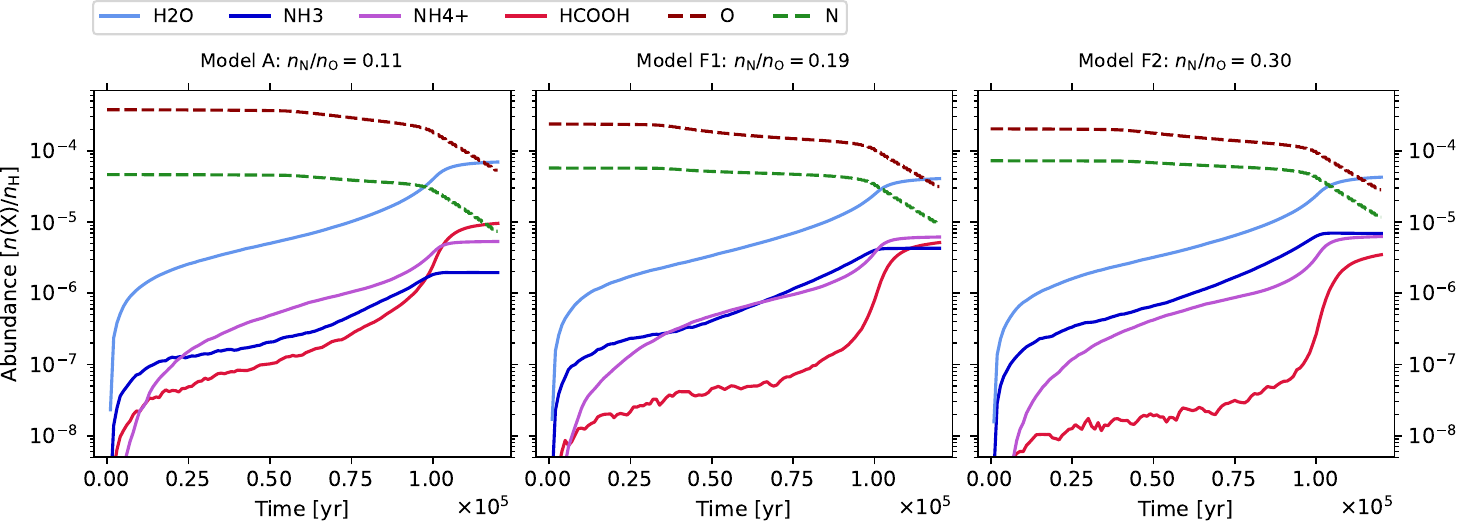}
      \caption{Selection of gas-phase (dashed) and grain-surface (solid) abundance profiles for Models~A, F1, and~F2, ordered from left to right according to increasing elemental N/O ratio, as indicated in the panel titles. Model~A uses elemental abundances representative of very low depletion, while Models~F1 and~F2 adopt abundance sets corresponding to more strongly depleted conditions.}\label{FigVaryElemAbun}
\end{figure*}

Overall, these results indicate that the extended timescales of Models~E1 and~E2 can partially compensate for lower accretion fluxes, allowing efficient salt formation even at lower densities. However, as no destruction mechanisms for ammonium salts are included in our network, the continued accumulation of these species over longer timescales should be interpreted with caution.

\subsubsection{Elemental abundances}\label{subsecElemAbun}
In Model~A we have opted to use elemental abundances (EA) of \citet{Jenkins2009} corresponding to the minimally depleted case. 
To evaluate how our results depend on the adopted elemental abundances, we also consider two additional models, Models~F1 and~F2, which both assume higher levels of elemental depletion. As listed in Table~\ref{TabElemAbun}, Model~F1 adopts the maximally depleted abundances of \citet{Jenkins2009}, while Model~F2 adopts the high-metal abundance set (EA2) tabulated by \citet{Wakelam2008}. Since nitrogen remains minimally depleted, especially in comparison to other elements such as oxygen and carbon, models of stronger depletion exhibit larger elemental N/O and N/C ratios (also shown in Table~\ref{TabElemAbun}). 
These ratios can significantly influence the competition between reaction pathways for nitrogen- and oxygen-bearing species.
\begin{table}[h!]
\caption{Elemental abundances with respect to $n_\mathrm{H}$}\label{TabElemAbun}
\centering
\begin{tabular}{lccc}
\hline\hline
Species & Model A\tablefootmark{a} & Model F1\tablefootmark{b} & Model F2\tablefootmark{c}  \\
\hline
He  & $0.1$ & $0.1$ & $0.09$ \\
N   & $6.18\times 10^{-5}$  & $6.18\times 10^{-5}$  & $7.60\times 10^{-5}$ \\ 
O   & $5.62\times 10^{-4}$  & $3.34\times 10^{-4}$  & $2.56\times 10^{4}$  \\
C   & $2.23\times 10^{-4}$  & $1.77\times 10^{-4}$  & $1.20\times 10^{-4}$  \\
Si  & $2.44\times 10^{-5}$  & $2.98\times 10^{-6}$  & $1.70\times 10^{-6}$ \\
Fe  & $3.88\times 10^{-6}$  & $2.01\times 10^{-7}$  & $2.00\times 10^{-7}$ \\
Mg  & $2.239\times 10^{-5}$ & $2.254\times 10^{-6}$ & $2.40\times 10^{-6}$ \\
\hline
$n_{\ce{N}}/n_{\ce{O}}$ & 0.11 & 0.19 & 0.30 \\
$n_{\ce{N}}/n_{\ce{C}}$ & 0.28 & 0.35 & 0.63 \\
\hline
\end{tabular}
\tablefoot{
\tablefoottext{a}{Minimally depleted case \citealp{Jenkins2009}.}
\tablefoottext{b}{Maximally depleted case \citealp{Jenkins2009}}.
\tablefoottext{c}{High-metal abundance (EA2) \citealp{Wakelam2008}.}
}
\end{table}

The abundance profiles throughout Phases~1 and~2 of the gas-phase atomic nitrogen and oxygen, together with the ice abundances of \ce{NH4+}, HCOOH, \ce{NH3} and \ce{H2O} are shown in Fig.~\ref{FigVaryElemAbun} for Models~A, F1, and~F2. These models are ordered according to increasing elemental N/O ratio. While only a modest increase in gas-phase atomic N and a slight decrease in atomic O are observed for Models~F1 and~F2 compared to Model~A, the elemental N/O ratio increases by almost a factor of two and three, respectively. 
This change has very little effect on the \ce{H2O} abundance and only a minor effect on the \ce{NH4+} abundance. However, it causes a clear transition in the relative abundances of HCOOH and \ce{NH3}. In Model~A, the final HCOOH abundance substantially exceeds that of \ce{NH3}, while both species reach comparable abundances in Model~F1. In Model~F2, \ce{NH3} becomes much more abundant than HCOOH. This transition is explained by the sensitivity of the HCOOH and \ce{NH3} formation pathways to the elemental N/O ratio. The pathways leading to these species compete for available atomic H, with increasing N/O favouring nitrogen hydrogenation over the oxygen- and carbon-bearing chemistry involved in HCOOH formation. Because the increased availability of one precursor of \ce{NH4+} (\ce{NH3}) is accompanied by a reduced abundance of the other (HCOOH), the \ce{NH4+} abundance remains robust across the different elemental abundance models considered here.

\subsection{Sensitivity of \ce{NH4+} formation to model variations}\label{subsecNH4+formation}
Section~\ref{subsecIceEvo} discussed the ice evolution of our main model, Model~A, whereas Sects.~\ref{subsecVaryChem} and~\ref{subsecVaryPhys} considered different chemical or physical variations of Model~A.
Here, we synthesize the results from these variations in order to assess how sensitive the formation of \ce{NH4+HCOO-} and \ce{NH4+OCN-}, as well as those of the major ice constituents, are to changes in the adopted model assumptions. We only include those models that have the correct two-step \ce{CO2} formation route via HOCO, i.e., all models except Model~B. 

\begin{figure*}
\sidecaption
\includegraphics[width=12.97cm]{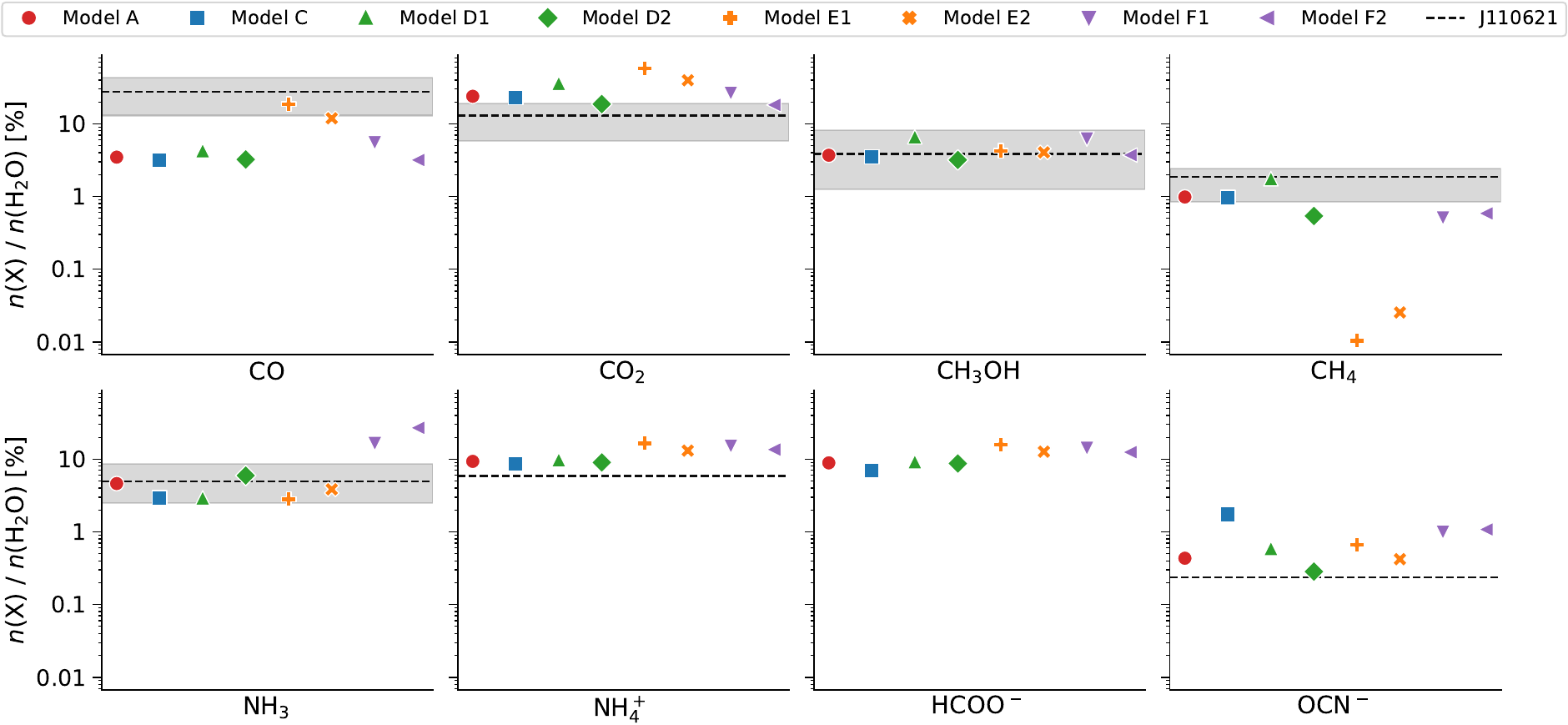}
\caption{Comparison of relative abundances of main ice species (top panels), \ce{NH3}, and \ce{NH4+} salts (bottom panels) for Model~A, its variations discussed in Sects.~\ref{subsecVaryHNCO}-\ref{subsecElemAbun}, and observations. Abundances are percentages relative to \ce{H2O}, taken at the onset of CO freeze-out, $t_\mathrm{freeze}$ (discussed in Sect.~\ref{subsecSlowerCollapse}). Colors indicate Model~A (red) and variations: \ce{CO + NH \to HNCO} (blue), cosmic-ray ionization rate (green), slower collapse (orange), and higher-depletion elemental abundances (purple). Grey dashed lines and shaded areas indicate observed values and $3\sigma$ confidence intervals, respectively, for sightline J110621 from \citet{McClure2023} where available.}\label{FigFinFracAbun}
\end{figure*}

An overview of the ice composition of Model~A and the different model variations is provided in Fig.~\ref{FigFinFracAbun}, which shows the fractional abundances relative to water for some of the main ice species (\ce{CO}, \ce{CO2}, \ce{CH3OH}, and \ce{CH4}), along with \ce{NH3} and the salt species \ce{NH4+}, \ce{HCOO-}, and \ce{OCN-}. For comparison, the observed abundances from \citet{McClure2023} towards background star J110621 in the Chamaeleon~I cloud -- which we regard as representative of a typical molecular-cloud environment and on which the parameter values for Model~A were loosely based -- are also indicated where available. Assessment of the CO column density indicates that the line of sight towards J110621 probes a source in an early stage of ice evolution, before catastrophic deposition of CO onto the grains has occurred \citep{McClure2023}. We hence indicate the abundances at $t=t_\text{freeze}$.

Figure~\ref{FigFinFracAbun} shows that all models have rather similar ice abundances with respect to \ce{H2O}, which are in  reasonable agreement with the observed values. The largest deviations from Model~A are observed for Models~E1 and~E2 and Models~F1 and~F2.
Models~E1 and~E2 exhibit higher abundances of \ce{CO} and \ce{CO2} and a much lower abundance of \ce{CH4}, which can be explained by their earlier conversion from C to CO in the gas phase.
Models~F1 and~F2 deviate mostly in N-bearing species, which are more dominant due to the higher N/O ratio in the gas. Finally, \ce{OCN-} is significantly more important in Model~C, because of the inclusion of the reaction \ce{CO + NH \to HNCO}.

Compared to observations, we find an overly efficient conversion of CO to \ce{CO2} in all our models, which generally results in an underestimation of the abundance of \ce{CO} and an overestimation of that of \ce{CO2}. 
Increased production of \ce{CO2} has also previously been encountered in other astrochemical models \citep{Garrod2011,Clement2023,Jimenez2025}, with different underlying mechanisms proposed to explain this behaviour. 
Although a detailed investigation of these mechanisms falls outside the scope of this work, we note that our KMC model includes \ce{CO2} formation through the radical HOCO. As discussed in Sect.~\ref{subsecEffectCO2}, formation of \ce{CO2} through the HOCO intermediate results in more modest \ce{CO2} production than allowing \ce{CO2} to form directly through reaction~\ref{reacCO2direct}, and therefore results in better agreement with the observed abundances presented here. Further improvement may nevertheless be possible.
Different conformers of HOCO may need to be considered explicitly, as detailed computational studies have shown the branching between the products of \ce{HOCO + H} to be conformer dependent, with \ce{CO2} formation being favoured by cis-HOCO \citep{Molpeceres2025}. The relative abundance of the different conformers may therefore affect the overall rate of \ce{CO2} formation and provide a further improvement to the treatment of the CO-to-\ce{CO2} balance.

Considering the ammonium salts and ammonia in Fig.~\ref{FigFinFracAbun}, we find that these species remain abundant in the ices across all model variations. The very similar fractional abundance of \ce{NH4+}, close to $10\%$ in all models, indicates that its formation is relatively insensitive to the range of physical and chemical conditions explored here. As salt production is dominated by ammonium formate among the two ammonium salts, the \ce{HCOO-} abundance closely follows that of \ce{NH4+} in all models. As discussed above, more variation in the fractional abundance is observed for \ce{OCN-}. 
Model~A and Models~D1 to~E2 show \ce{NH4+}, \ce{OCN-}, and \ce{NH3} abundances in good agreement with the observed values. 
In general, we find a slightly higher \ce{NH4+}-to-\ce{NH3} ratio than observed. This balance may be affected by the absence of dissociation reactions for the ammonium salts in our model. However, photodissociation rates under astrophysically relevant ice conditions for \ce{NH4+HCOO-} and \ce{NH4+OCN-}, and for ammonium salts in general, remain unconstrained. Moreover, the ionic environment of the ice may lead to rapid recombination or proton-transfer reactions following ionization, making it unclear whether salt destruction in the ice can be adequately represented by a simple dissociation reaction. Energetic-particle processing may therefore provide a more appropriate framework for investigating the destruction and reprocessing of ammonium salts. Proton-irradiation experiments on \ce{NH4+SH-}, for example, show that irradiation leads to the formation of \ce{NH3} and reactive sulfur-bearing species, demonstrating that ionization can initiate subsequent chemistry involving both the ammonium ion and its counterion \cite{Loeffler2015}. Thus, rather than introducing an isolated dissociation reaction for the ammonium salts, a more complete treatment would involve an ionization-driven reaction network in which the products of salt processing can undergo subsequent proton-transfer, recombination, and other ion-molecule reactions.

Furthermore, additional types of ammonium salts could be included, as a wider range of counterions is expected to be present in the ice. Plausible candidates include sulfur-bearing ionic species, in particular \ce{NH4+SH-} \citep{Vitorino2024,Slavicinska2025},  \ce{CN-} \citep{Noble2012}, and \ce{Cl-} \citep{Altwegg2022}. 
Since our models show that ammonium salt formation is limited by the availability of \ce{NH3}, we expect that the inclusion of additional counterions and their associated acids, would not substantially increase the total \ce{NH4+} abundance. Inclusion of other \ce{NH4+} salts in our models would simply redistribute ammonium among the different anions, while leaving the ammonia reservoir (\ce{NH3} and \ce{NH4+} abundance) relatively unchanged and in reasonable agreement with observation.

The contribution to ammonia production by the molecule forming in the gas and subsequently accreting onto the grain has been neglected in the present study, since we find that \ce{NH3} and the related species \ce{NH} and \ce{NH2} together contribute less than 0.015\%  to the elemental N budget throughout the course of the simulation.

The dust temperature was kept constant at $T=10$~K in all models and was therefore not explored as a variable. Whereas diffusion and desorption of hydrogen is found to respond sensitively on the dust temperature, KMC simulations of ice chemistry that are mainly driven by radical-radical reactions show only a minor temperature dependence for the ice abundances \citep{Jimenez2025,Simons2020}. We reason that this is because \emph{(i)} H atoms find a reactant before desorption independent of the temperature in the 10-15~K range and \emph{(ii)} most reactions are non-diffusive radical-radical reactions that are rather temperature independent. 
Although the rate constant of the acid-base reactions depends exponentially on the temperature, we do not expect a strong temperature dependence of the \ce{NH4+} abundance either, since the acid-base reactions are \ce{NH3}-limited and not reaction-barrier limited.

The robust formation of the two ammonium salts and the limited variation in the fractional abundance of \ce{NH4+} across the different models can be attributed to several factors. First, the main change in several variations of Model~A is the abundance of atomic H. Although this affects the absolute amount of \ce{NH4+} formed, it also affects the formation of \ce{H2O} in a similar and proportional manner, leaving the ratio \ce{NH4+} / \ce{H2O} relatively unchanged. Second, we find that changes in the abundance of one precursor to the ammonium salts can be partially compensated by changes in the abundance of the other (e.g. HCOOH and \ce{NH3}), as their formation pathways compete for the available atomic H on the grains.

\section{Conclusions}\label{secConclusions}
Using our microscopic KMC model of grain-surface chemistry, combined with gas-phase and physical cloud evolution calculations performed with UCLCHEM, we have modeled the formation of ammonium formate (\ce{NH4+HCOO-}) and ammonium cyanate (\ce{NH4+OCN-}) in interstellar ice, motivated by the JWST detection of \ce{NH4+} toward the Chamaeleon~I dark molecular cloud. This model framework allows us to treat the formation of \ce{NH4+} salts in the ice in detail while following the chemical and physical evolution of a collapsing molecular cloud. 
\begin{enumerate}
    \item The abundances of \ce{NH4+}, \ce{HCOO-} and, to a lesser extent, \ce{OCN-} are found to be remarkably robust across all model variations explored in this work, particularly when expressed with respect to \ce{H2O}. Moreover, the relative abundances of \ce{NH4+} and \ce{OCN-} are in good agreement with observations.
    
    \item Salt production is dominated by \ce{NH4+HCOO-}, which is predominantly formed in the \ce{H2O}-rich ice mantle. \ce{NH4+OCN-} is mainly formed at a later stage, within the interface between the \ce{H2O}- and CO-rich ice, when the formation of its precursor acid HNCO is enhanced. The production of both salts is ultimately limited by the availability of \ce{NH3}.
    
    \item An increase in the cosmic-ray ionization rate has a limited effect on salt production with respect to \ce{H2O}. The higher dissociation rate is compensated by the higher production rate because of the increased H abundance. This results in a higher absolute abundance of both \ce{NH4+} and \ce{H2O}.
\end{enumerate}

Given the robust nature of \ce{NH4+} production, we expect that adding further salt species -- such as ammonium sulfur salts -- would primarily redistribute \ce{NH4+} among different counter anions without significantly altering its total abundance. To further improve the chemical precision of our results, future extensions of the network should incorporate salt destruction processes (e.g., dissociation by energetic particles) to better balance the \ce{NH4+}/\ce{NH3} ratio, constrain the branching ratios of the \ce{HOCO + H} reaction, and implement environment-dependent rate coefficients. These extensions will allow for a more nuanced treatment of the evolution of ammonium salts in interstellar ices.

%

 \bibliographystyle{bibtex/aa.bst} 
\begin{appendix}

\section{\ce{H2} formation efficiency}\label{AppenH2formationEfficiency}
The default UCLCHEM rate for \ce{H2} formation on grains follows the treatment of \citet{Cazaux2002,Cazaux2004} and is given by
\begin{align}\label{eq:RH2form}
R_{\ce{H2},\mathrm{form}} = \frac{1}{2}\epsilon\,S_\mathrm{H}\,\sigma\,v_\mathrm{th}\,n(\mathrm{H})\,n_d,
\end{align}
where $\sigma$ is the grain cross-section, $v_\mathrm{th}$ the thermal velocity, and $n(\mathrm{H})$ and $n_d$ number density of H atoms in the gas and of dust grains, respectively. $S_\mathrm{H}$ denotes the sticking probability of an H atom onto a grain and is temperature dependent following \citet{Hollenbach1979}. The recombination efficiency, $\epsilon$, is calculated according to \citet{Cazaux2002} and within the UCLCHEM simulations of the collapsing molecular cloud (Phase~1 and~2; see Section~\ref{subsecPhysicalParameters}), this factor assumes a constant value relatively close to the maximum efficiency limit of $1$, namely $\epsilon=0.83$. 

\begin{figure}
    \centering
    \includegraphics[width=1.0\hsize]{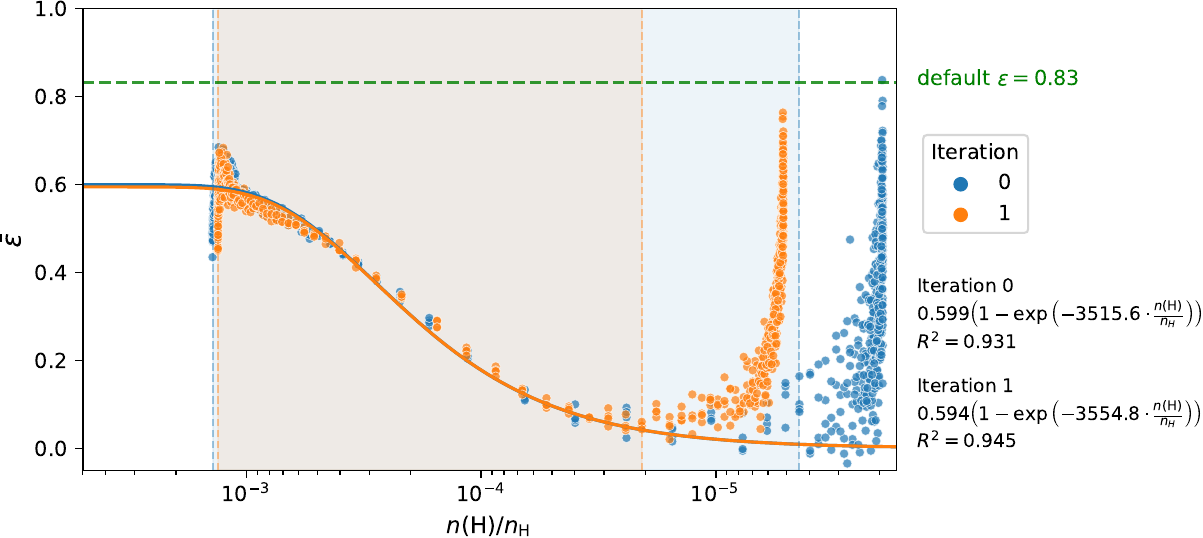}
    \caption{Determination of the custom efficiency factor, $\bar{\epsilon}_\mathrm{fit}(n(\mathrm{H}))$, by which the default \ce{H2}-recombination efficiency, $\epsilon$, in UCLCHEM is replaced. The value of the latter is depicted by the green dashed line. Blue and orange data points represent the efficiency determined using Eq.~(\ref{eq:KMC_H2efficiency}) from the KMC coupled to the UCLCHEM simulation with the default (iteration 0) and the custom (iteration 1) efficiency respectively. Data points within the selected range (blue- and orange-shaded area) are fit by an exponential growth function (Eq.~(\ref{eq:fit_H2efficiency})) yielding $\bar{\epsilon}_\mathrm{fit}(n(\mathrm{H}))$ (for iteration 0 and 1 respectively). The KMC simulations are iterations of Model~A, as described in Sect.~\ref{subsecIceEvo}. The datapoints were obtained by averaging over $1000$ year intervals and 4 independent runs.}\label{fig:H2efficiency}
\end{figure}

We can compare the efficiency factor, $\epsilon$, to the actual efficiency encountered in the KMC simulations, which we denote as $\bar{\epsilon}$. In the simulations, \ce{H2} that desorbs can either originate from \ce{H2} formation or from desorption of adsorbed \ce{H2} from the gas phase. This is accounted for in the following way:
\begin{align}\label{eq:KMC_H2efficiency}
\bar{\epsilon} = 2\frac{N_{\ce{H2},\mathrm{des}} - N_{\ce{H2},\mathrm{dep}}}{N_{\mathrm{H},\mathrm{dep}}},
\end{align}
with $N_{\ce{H2},\mathrm{des}}$ the number of desorbed \ce{H2} molecules within a certain time bin, and $N_{\mathrm{X},\mathrm{dep}}$ is the number of species X (= H or \ce{H2}) landing on the grain within the same time bin. Fig.~\ref{fig:H2efficiency} shows the efficiency calculated from the corresponding KMC simulation using Eq.~(\ref{eq:KMC_H2efficiency}) (iteration~0, blue data points) as a function of the fractional abundance of atomic H in the gas. In contrast to the default value, the resulting efficiency is found to depend strongly on the atomic H gas abundance. As the atomic H abundance decreases -- which, in practice, corresponds to the temporal evolution of the ice mantle -- $\bar{\epsilon}$ initially lies below the default value, with $\bar{\epsilon}\sim 0.6$ at $n(\mathrm{H})/n_\mathrm{H}\sim 10^{-3}$, and subsequently decreases towards zero for $n(\mathrm{H})/n_\mathrm{H}\sim 10^{-5}$. We attribute these lower values of $\bar{\epsilon}$ to the consumption of atomic H in surface reactions other than \ce{H2} recombination. The abundance ratio $n(\mathrm{H})/n_\mathrm{H}\sim10^{-5}$ corresponds to the stage at which CO ice begins to form on top of the already established \ce{H2O} ice mantle. The interface between these two ice layers is also where the chemical activity is expected to peak. Indeed, a minimum in $\bar{\epsilon}$ can be observed around $n(\mathrm{H})/n_\mathrm{H}<10^{-5}$, after which it increases again towards the default value of $\sim 0.83$. This latter regime corresponds to a later evolutionary stage in which most reactive atomic species have already been depleted, and H atoms are mostly involved in subsequent hydrogenation and abstraction reactions leading to efficient \ce{H2}-formation cycles.
\begin{figure}
    \centering
    \includegraphics[width=\hsize]{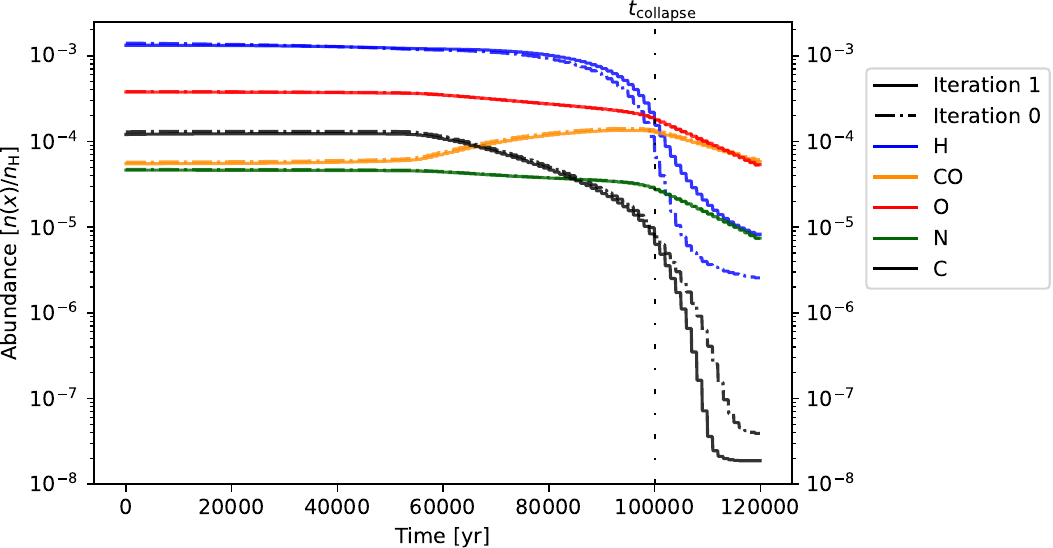}
    \caption{Time evolution of the gas abundances of the main atomic species and CO during the collapse and dense-cloud stages (Phases~1 and~2) for iterations~0 and~1, corresponding to the default and custom \ce{H2} recombination efficiencies, respectively. The use of the custom efficiency primarily affects the abundance of atomic H, while the abundances of O, N, and CO remain largely unchanged.}\label{fig:H2effGasAbun}
\end{figure}

To better reproduce the \ce{H2}-formation efficiency encountered in the KMC simulation, we replace the default recombination efficiency from\ \citet{Cazaux2002} with a custom efficiency factor ($\bar{\epsilon}_{\mathrm{fit}}$). To this end, we first fit the efficiencies obtained from the KMC simulation which was still coupled to the default UCLCHEM treatment (iteration~0) with an asymptotic exponential function
\begin{align}\label{eq:fit_H2efficiency}
\bar{\epsilon}_\mathrm{fit}\left(n(\mathrm{H})\right)
= a\left[1-\exp\left(-b\frac{n(\mathrm{H})}{n_\mathrm{H}}\right)\right],
\end{align}
where $a$ denotes the asymptotic maximum efficiency and $b$ determines the exponential scaling. Both quantities are treated as free parameters. Only a subset of the KMC data points are included in the fit, corresponding to the fractional atomic H abundance range $1.3\times10^{-3} \ge n(\mathrm{H})/n_\mathrm{H} \ge 4.5\times10^{-6}$ (blue shaded area). At lower atomic H abundances, the efficiency increases rapidly and cannot be reproduced by the adopted functional form. This regime roughly corresponds to the end of phase~2 and beyond (Sect.~\ref{secMethods}), and as mentioned above, is a phase where the ice chemistry has largely settled. This regime is not important for the purposes of the present study and therefore we allow for the underestimation of surface-\ce{H2} formation there. Optimizing the fit yields $a=0.6$ and $b=3500$.

Next, we perform a second UCLCHEM simulation replacing $\epsilon$ in Eq.~(\ref{eq:RH2form}) by $\bar{\epsilon}_\text{fit}$ and using a sticking probability of H atoms equal to unity, as relatively recent studies suggest that this is a good approximation for gas and grain temperatures of $10$ K \citep{Veeraghattam2014,Dupuy2016}. The efficiencies extracted from the corresponding KMC simulation (iteration~1) are also shown in Fig.~\ref{fig:H2efficiency}. Repeating the fitting procedure yields nearly identical values of $a$ and $b$, demonstrating that $\bar{\epsilon}_\mathrm{fit}(n(\mathrm{H}))$ converges after a single iteration. Again only a subset of the KMC data points are included in the fit ($1.3\times10^{-3} \ge n(\mathrm{H})/n_\mathrm{H} \ge 2\times10^{-5}$).

Employing the custom efficiency factor has a negligible effect on the grain-surface formation of \ce{H2} in the KMC simulations. It does, however, significantly affect the atomic hydrogen abundance predicted by UCLCHEM. Fig.~\ref{fig:H2effGasAbun} shows the gas abundances of the main atomic species, together with CO, during Phases~1 and~2. While the abundances of O, N, and CO remain essentially unchanged throughout the collapse and subsequent dense-cloud phase, the abundance of atomic H exhibits a less pronounced decline during Phase~2 in iteration~1 than in iteration~0. As a consequence, the iteration~1 KMC simulation experiences a higher deposition flux of atomic H, leading to an enhanced surface abundance of H atoms, particularly during Phase~2.

\begin{figure}
    \centering
    \includegraphics[width=\hsize]{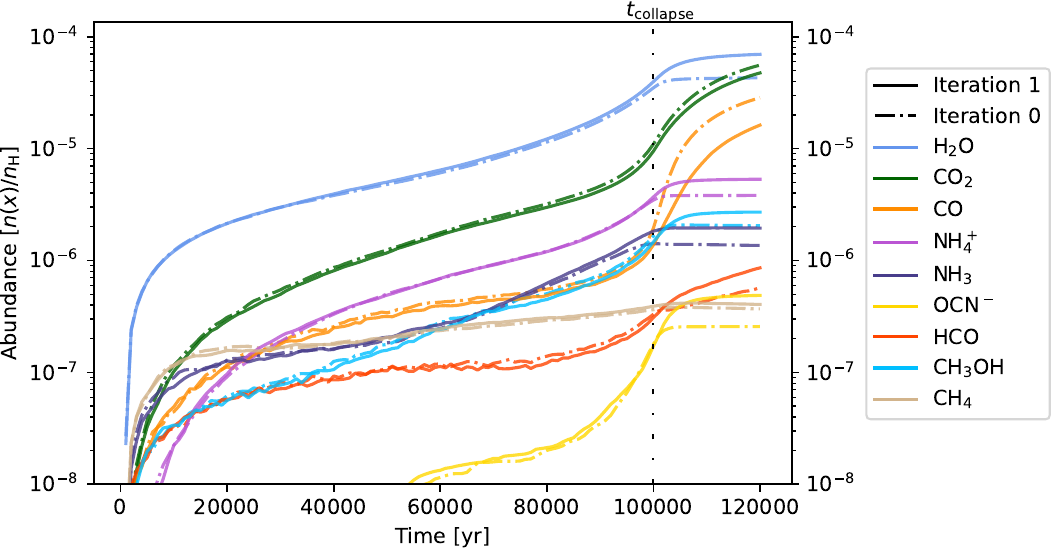}
    \caption{Time evolution of selected grain-surface species during the collapse and dense-cloud stages (Phases~1 and~2) for iterations~0 and~1, corresponding to the default and custom \ce{H2} recombination efficiencies, respectively. The increased atomic H abundance in iteration~1 enhances the formation of hydrides and CO-hydrogenation products during Phase~2. Consequently, larger abundances of ammonium salts are produced, although their formation still occurs predominantly in the \ce{H2O}-rich ice during the collapse phase for both iterations.}\label{fig:H2effAggAbun}
\end{figure}

The time evolution of the abundances of several ice species during phases~1 and~2 are shown in Fig.~\ref{fig:H2effAggAbun}. For iteration~1, we indeed observe enhanced formation of hydrides around and after the collapse time, as illustrated by the increased abundances of \ce{H2O} and \ce{NH3}. This enhancement is a direct consequence of the increased availability of atomic H on the grain surface.
\clearpage
\onecolumn
\section{Additional model details}\label{AppenModelDetails}
\begin{figure}[ht!]
   \centering
   \includegraphics[width=1.\textwidth]{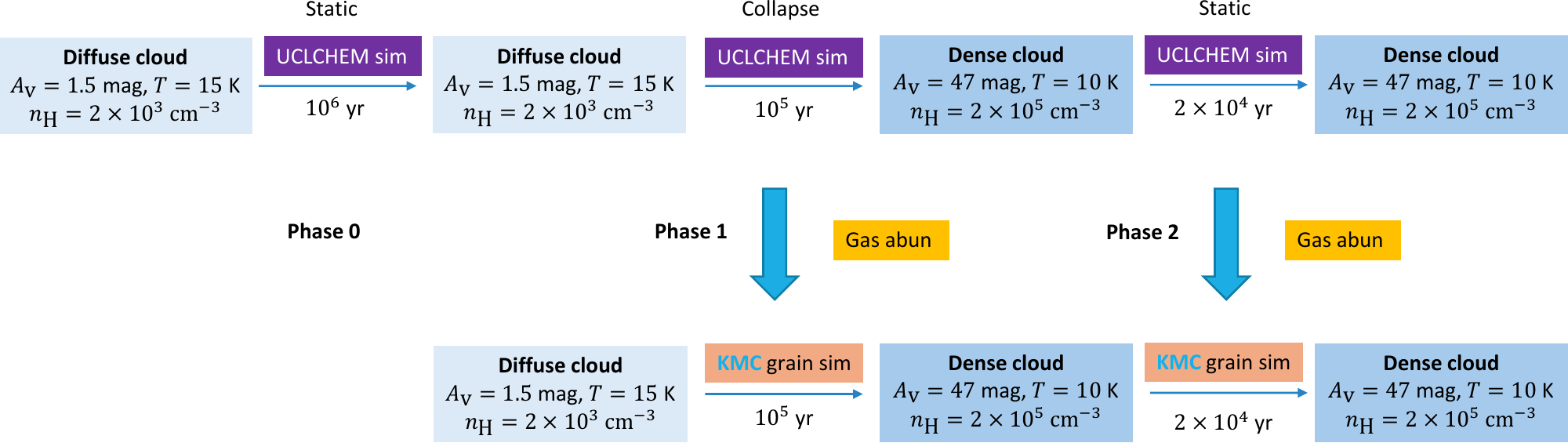}
      \caption{Flowchart of our setup simulating the molecular cloud with the UCLCHEM and the KMC code. Parameter values shown for modeling  line of sight towards J110621 of the Chamaeleon I cloud, which we take as an example of a typical dense cloud.}\label{FigSimSchematic}
\end{figure}

\begin{table}[h!]
\caption{Overview of newly added reactions.}\label{TabAddedReactions}
\label{tab:added_reactions}
\centering
\begin{tabular}{lclcc>{\raggedright\arraybackslash}p{6.4cm}}
\hline\hline
Reaction & & Products & Rate [s$^{-1}$] & Branching ratio & Ref. \\
\hline
\ce{NH3 + HCOOH} & $\rightarrow$ & \ce{NH4+ HCOO-} & $1.28\times10^{-6}$ & 1    & \citealp{Bergner2016} \\
\ce{NH3 + HNCO}  & $\rightarrow$ & \ce{NH4+ OCN-}  & $2.88\times10^{-5}$ & 1    & \citealp{Mispelaer2012} \\
\ce{HCO + OH}    & $\rightarrow$ & \ce{HCOOH}      & $2\times10^{11}$    & 0.5  & \citealp{Dontgen2016} \\
\ce{HCO + OH}    & $\rightarrow$ & \ce{CO + H2O}   & $2\times10^{11}$    & 0.5  & \citealp{Dontgen2016} \\
\ce{CH2OH + O}   & $\rightarrow$ & \ce{HCOOH + H}  & $2\times10^{11}$    & 0.5  & \citealp{Seetula1994} \\
\ce{CH2OH + O}   & $\rightarrow$ & \ce{H2CO + OH}  & $2\times10^{11}$    & 0.5  & \citealp{Seetula1994} \\
\ce{CO + OH}     & $\rightarrow$ & \ce{HOCO}       & $1\times10^{10}$    & 1    & \citealp{Molpeceres2023} \\
\ce{HOCO + H}    & $\rightarrow$ & \ce{HCOOH}      & $2\times10^{11}$    & 0.33 & \\
\ce{HOCO + H}    & $\rightarrow$ & \ce{CO2 + H2}   & $2\times10^{11}$    & 0.33 & \\
\ce{HOCO + H}    & $\rightarrow$ & \ce{CO + H2O}   & $2\times10^{11}$    & 0.33 & \\
\ce{HOCO + N}    & $\rightarrow$ & \ce{HOCON}      & $2\times10^{11}$    & 0.8  & \citealp{Dibble2010} \\
\ce{HOCON + H}   & $\rightarrow$ & \ce{OCN + H2O}  & $2\times10^{11}$    & 0.3  & \\
\ce{HOCON + H}   & $\rightarrow$ & \ce{HNCO + OH}  & $2\times10^{11}$    & 0.3  & \\
\ce{OCN + H}     & $\rightarrow$ & \ce{HNCO}       & $2\times10^{11}$    & 1    & \\
\ce{CO + O}      & $\rightarrow$ & \ce{CO2}        & $1\times10^{-10}$   & 1    & \citealp{Goumans2008} \\
\ce{HCO + O}     & $\rightarrow$ & \ce{CO2 + H}    & $2\times10^{11}$    & 0.5  & \citealp{Baulch2005,Campbell1978,Westenberg1972} \\
\ce{HCO + O}     & $\rightarrow$ & \ce{CO + OH}    & $2\times10^{11}$    & 0.5  & \citealp{Baulch2005,Campbell1978,Westenberg1972} \\
\hline
\end{tabular}
\end{table}






\FloatBarrier 
\clearpage
\end{appendix}

\end{document}